\pdfoutput=1
\documentclass[aps,10pt,prc,tightenlines,twocolumn,twoside,showpacs,superscriptaddress,nolongbibliography]{revtex4-2}
\usepackage[caption=false]{subfig}

\usepackage[hidelinks]{hyperref}
\usepackage{color}
\usepackage{graphicx}
\usepackage{siunitx}
\usepackage{amsmath,amssymb,amsfonts}
\usepackage{bm}
\usepackage{lineno}

\usepackage{tensor}
\usepackage{changepage}   
\usepackage{slashed}
\usepackage{epstopdf-base}

\usepackage[normalem]{ulem}  

\newcommand\ddfrac[2]{\frac{\displaystyle #1}{\displaystyle #2}}

\def \be {\begin{equation} }
	\def \ee {\end{equation}}
\def \bes {\begin{subequations} }
	\def \ees {\end{subequations}}

\def \no {\nonumber}
\def \a {\alpha}
\def \b {\beta}

\def \d {\delta}
\def \e {\epsilon}
\def \g {\gamma}

\def \o {\omega}
\def \p {\psi}
\def \r {\rho}
\def \l {\lambda}
\def \m {\mu}
\def \n {\nu}
\def \s {\sigma}
\def \t {\tau}

\def \La{\Lambda}

\def \pd {\partial}

\def \<{\langle}
\def \>{\rangle}
\def \+{\dagger}

\def \[{\left[}
\def \]{\right]}

\def \vp {\bm{p}}
\def \vq {\bm{q}}
\def \vu {\bm{u}}
\def \vx {\bm{x}}
\def \vy {\bm{y}}

\def \vk {\bm{k}}

\def \ve {\varepsilon}
\def \hn {\hat{n}}

\def \D {\Delta}

\def \<{\langle}
\def \>{\rangle}
\def \+{\dagger}

\def \[{\left[}
\def \]{\right]}

\def \sS {{\cal S}}
\def \sT {{\cal T}}
\def \srho 
{{\tilde \rho}}
\def \Tr {\text{Tr}}

\renewcommand{\sout}{\bgroup \color{red} \ULdepth=-.5ex \ULset}

\begin{document}
	\title{Zubarev response approach to polarization phenomena in local equilibrium}
	
	\author{Youyu Li
	}
	\affiliation{School of Physics and Electronics, Hunan University, Changsha 410082, China} 
	\affiliation{Hunan Provincial Key Laboratory of High-Energy Scale Physics
		and Applications, Hunan University, Changsha 410082, China}
	\affiliation{School of Physical Science and Technology, Lanzhou University, Lanzhou 730000, China}
	\author{Shuai Y.\,F.~Liu
	}
	\email[Corresponding author.\\]{lshphy@hnu.edu.cn}
	\affiliation{School of Physics and Electronics, Hunan University, Changsha 410082, China} 
	\affiliation{Hunan Provincial Key Laboratory of High-Energy Scale Physics
		and Applications, Hunan University, Changsha 410082, China}

	\date{\today}
\begin{abstract}
Using the expansion of Zubarev's density operator, we develop a linear response approach to study various spin physics in a locally equilibrated medium, particularly focusing on various polarization phenomena  in heavy-ion collisions. Specifically, we connect familiar correlation functions and diagrammatic methods to the Zubarev formalism, enabling the use of established techniques like the Matsubara/imaginary time formalism to facilitate calculations.
For a spin-1/2 particle, we re-derive its vector polarization using this Zubarev response approach, which exactly reproduces with our previous results based on Luttinger's method. For a spin-1 particle, we calculate the vector polarization and find the expected contributions from vorticity, temperature gradients, and shear, which are identical to those for spin-1/2 particles except for a factor of 4/3 as expected. For the tensor polarization and spin alignment of a spin-1 boson, we explicitly prove that the non-dissipative contribution is zero at leading order in gradients, and briefly reiterate our previous findings for the dissipative contribution with further discussions on several concerns.
Additionally, we discuss several relevant subtleties and questions, including an alternative derivation for Zubarev response approach, the covariance issues of different spin density matrix definitions, a further explanation of slow and fast modes, the mode selection scheme, etc. We also discuss skeleton expansions, higher-order contributions, and non-perturbative methods, particularly their potential connection to lattice field theory.
In summary, this work discusses the foundations and subtleties of Zubarev response approach, with specific examples from spin physics in heavy-ion collisions.
\end{abstract}
\maketitle
\flushbottom
\section{Introduction}
Following the pioneering theoretical~\cite{Liang:2004ph} and experimental breakthroughs~\cite{STAR:2017ckg} in the study of the Lambda baryon’s global polarization, spin-related observables have emerged as a promising avenue for exploring the physics of hot and dense matter created in heavy-ion collisions. In recent years, new observables, such as local polarization along the beam direction~\cite{STAR:2018gyt, ALICE:2021pzu} and spin alignment of vector mesons~\cite{ALICE:2019aid, STAR:2022fan}, have been successfully measured, leading to the discovery of intriguing polarization phenomena. Along with the rapid advancement of theory~\cite{Liang:2004xn,Li:2017slc, Karpenko:2016jyx,Becattini:2017gcx,Becattini:2018duy,Liu:2019krs,Li:2020eon,Speranza:2020ilk,Fukushima:2020ucl,Hongo:2021ona,Liu:2021uhn,Fu:2021pok,Becattini:2021suc,Becattini:2021iol,Li:2022vmb,Wagner:2022gza,Sheng:2022wsy,Wang:2020pej,Ivanov:2020qqe,Lisa:2021zkj,Florkowski:2021wvk,Florkowski:2021xvy,Yi:2021ryh,Muller:2021hpe,Lin:2022tma,Kumar:2022ylt,Wu:2022mkr,Fang:2023bbw,Kumar:2023ojl,Hidaka:2023oze,Dong:2023cng,Gao:2023wwo,Wang:2024lis,Sheng:2024pbw,Wang:2024lis,Weickgenannt:2024esg}, our understanding of relativistic spin polarization phenomena has significantly deepened in recent years, making spin-related physics a prominent frontier in heavy-ion collisions.

Among the various spin/polarization physics, the spin (vector) polarization of spin-1/2 particles has attracted the most attention,  since the first experimental observable was the global polarization of the spin-1/2 Lambda hyperon~\cite{Adam:2018ivw}. Theoretical predictions~\cite{Li:2017slc, Karpenko:2016jyx} based on a thermal vorticity formula~\cite{Becattini:2013fla} agree with experimental observations~\cite{Adam:2018ivw}, partly confirming the success of  ``thermal vorticity polarization" picture. However, subsequent measurements of more differential observables, such as local Lambda polarization~\cite{Niida:2018hfw, Adam:2019srw, ALICE:2021pzu}, reveal a sign opposite to that predicted by this model~\cite{Becattini:2017gcx}, leading to the so-called ``spin sign puzzle" in heavy-ion physics. Further theoretical developments identified a new mechanism for generating local polarization—shear-induced polarization~\cite{Liu:2021uhn, Fu:2021pok, Becattini:2021suc,Becattini:2021iol}—which make important progress on solving this ``sign" problem. Theory and experiment evolve together, driving rapid advancement of our understanding on the polarization phenomena of spin-1/2 particle  in heavy-ion collisions.

On the other hand, although vector polarization also exists for spin-1 vector mesons, research in this area is relatively limited due to a lack of experimental data. Theoretically, we have briefly discussed in Ref.~\cite{Li:2022vmb} that the polarization for spin-1 mesons is 4/3 of that for spin-1/2 particles, where shear-induced polarization also plays a role. Meanwhile, few other studies have also explored this topic in various contexts\cite{Hattori:2020gqh, Wagner:2022gza}. For better understanding of general polarization phenomena in local equilibrium and in heavy-ion collisions, further investigation into this subject would be valuable.

However, for vector mesons, another type of polarization—tensor polarization/spin alignment—has been observed in experiments~\cite{STAR:2008lcm, ALICE:2019aid, STAR:2022fan, ALICE:2022sli} for particles such as $K^*$, $\phi$, and $J/\psi$. The magnitude of spin alignment, however, is surprisingly large,  orders of magnitude lager than the predictions in Ref.\cite{Becattini:2007nd, Becattini:2007sr, Xia:2020tyd, Becattini:2022zvf}, since the contributions in these works are of higher order in gradients. Moreover, the observed dependence on transverse momentum ($p_T$), centrality, and beam energy, along with intriguing sign-flipping behavior, presents a challenge for current theoretical models. Recently, several  ideas have been proposed to address this puzzle~\cite{Sheng:2019kmk, Muller2021, Sheng:2022ffb, Sheng:2022wsy}. However, the leading-order thermal force effects, i.e., leading-order gradient contributions, have not been fully explored, as dissipative effects have largely been overlooked in most of the studies. The presence of dissipative contributions to spin alignment was first identified in our work~\cite{Li:2022vmb} and later confirmed by Ref.\cite{Wagner:2022gza, Dong:2023cng}. As shown in Ref.\cite{Li:2022vmb}, these dissipative contributions can potentially generate large spin alignment with dependencies on $p_T$, centrality, and beam energy that can  resemble the trends observed in experiments.

As demonstrated in our previous work~\cite{Liu:2021uhn, Li:2022vmb}, linear response theory is a powerful tool for understanding how thermal forces, such as hydrodynamic gradients, affect polarization phenomena near local equilibrium. However, the theoretical frameworks in Ref.\cite{Liu:2021uhn} and Ref.\cite{Li:2022vmb} are different. The former is based on Luttinger's mechanical perspective~\cite{Luttinger1964} on linear response, while the latter is based on Zubarev approach~\cite{zubarev1974nonequilibrium}. Aiming to construct a unified theoretical framework, we develop a linear response theory based on Zubarev's formalism--a familiar theory in community--to systematically investigate both dissipative and non-dissipative contributions to vector and tensor polarization for spin-1/2 and spin-1 particles within a coherent framework.
In comparison with other efforts~\cite{Sheng:2024pbw, Zhang:2024mhs, Yang:2024fkn}, the approach developed here places particular emphasis on connecting Zubarev's formalism with modern quantum field theory (QFT) concepts and techniques, such as retarded/Euclidean time Green’s functions and diagrammatic techniques, which can simplify the calculations. Moreover, this connection provides a non-perturbative meaning of many quantities. Various non-perturbative approaches, such as the Dyson-Schwinger equation(DSE)~\cite{Roberts:2000aa}, functional renormalization group(FRG)~\cite{Fu:2019hdw}, lattice QFT, and the T-matrix approach~\cite{Liu:2017qah, Liu:2016ysz}, can be employed to study the problem in principle.

The paper will cover many details and subtleties in the derivation and application of the Zubarev response approach along with the main track of the paper, and these materials will be organized as follows. 
In Sec.~\ref{sec_Zuba}, we briefly review the ideas, derivations, and subtleties of the Zubarev-based linear response approach, preparing it for subsequent application, while a alternative method of derivation is discussed in Sec.~\ref{sec_deco}. In Sec.~\ref{sec_spinhalf}, we re-derive the polarizations presented in \cite{Liu:2021uhn}, addressing subtleties related to slow and fast modes (Sec.~\ref{sec_mode}). In Sec.~\ref{sec_spinonevec}, we analyze the vector polarization of spin-1 particles, highlighting subtleties concerning the covariance problem of spin density matrices under various definitions (Sec.~\ref{sec_density}). In Sec.~\ref{sec_spinoneten}, we study tensor polarization and spin alignment, demonstrating the vanishing non-dissipative contribution at leading gradient order. A brief discussion of dissipative contributions is included, though the primary calculations are detailed in our previous work~\cite{Li:2022vmb}. Meanwhile, a discussion of skeleton expansions, higher-order effects, and non-perturbative methods is included in different context Sec.~\ref{sec_conti} and Sec.~\ref{sec_skeleton}. Finally, we summarize our findings in Sec.~\ref{sec_sum}.

\section{Zubarev response approach}
\label{sec_Zuba}
For completeness, we briefly review Zubarev's formalism following two references~\cite{zubarev1974nonequilibrium,Hosoya:1983id} for non-equilibrium physics. This formalism is primarily used to construct non-equilibrium transport equations. However, in this work, instead of deriving transport equations, we focus on developing the linear response approach based on Zubarev's density operator for a locally equilibrated hydrodynamic system, where we assume small flow velocities (in the rest frame of the medium) and small flow gradients.

The core of Zubarev's formalism is to construct a stationary density matrix in the Heisenberg picture that satisfies \(d\rho/dt = 0\), but in terms of time-dependent dynamical parameters. In the hydrodynamic scenario discussed here, we choose these parameters as \(\b^\mu(x) = \b(x) u^\mu(x)\), where \(\b(x)\) represents the local inverse temperature at a given spacetime point $x=(t,\vx)$, and \(u^\mu(x)\) is the flow velocity of the medium. Based on the maximum information entropy \(-\Tr[\rho \ln \rho]\) for a density operator $\rho$ with constraint \(\langle T^{0\nu} \rangle = \Tr[\rho T^{0\nu}]\)~\cite{zubarev1974nonequilibrium,Becattini:2018duy},  we obtain a local equilibrium density operator of the form $\propto\exp[-\int d^3x \, \beta_\mu T^{0\mu}]$, which is an intuitive extension of the global equilibrium density operator $\propto\exp[-\beta H]$. However, it is straightforward to see that \(d\rho/dt \neq 0\), meaning that it does not represent a physical density operator in the Heisenberg picture. To make it stationary, we define a time-averaged operator of the exponent \(\beta_\mu T^{0\mu}\)~\cite{zubarev1974nonequilibrium,Hosoya:1983id} as
\begin{align}
	\label{eq_B_op1}
	\mathcal{B}(t, \vx)=\epsilon\int^{t}_{-\infty} d t' e^{-\epsilon (t-t')}  \beta^\g (t', \vx)T_{0\g}(t', \vx) 
\end{align}
where the derivatives $d\mathcal{B}(t,\vx)/dt=0$ in small $\epsilon$ limit. Meanwhile, after performing a partial integration over time, a more useful expression is obtained as
\begin{align}
	\label{eq_B_op2}
	&\mathcal{B}(t,\vx)=\beta^\g (t, \vx)T_{0\g}(t, \vx)-\int_{-\infty}^{t} dt' e^{-\epsilon (t-t')} \no\\
	&\times\left[\pd^{0}\beta^{\g}(t', \vx)T_{0\g}(t', \vx)+\beta^{\g}(t', \vx)\pd^{0}T_{0\g}(t', \vx)\right]
\end{align}
The first term represents the expected local equilibrium density operator, which is at the same time $t$ with our operators at a given $t$. Therefore, there is no time evolution or dynamics when using the first term to trace the operator at a given $t$ , resulting in no dissipation in this term. On the other hand, all dynamics are contained in the second term, where dissipative contributions can be found. The \( e^{-\epsilon (t-t')} \) term becomes negligible if all quantities converge to zero properly in the infinite past. To simplify the notation, we make the \( e^{-\epsilon (t-t')} \) implicit; however, whenever convergence issues arise, we will reintroduce it. Considering the equation of motion for energy momentum tensor $\pd_\mu T^{\mu\nu}=0$ is $\pd^{0}T_{0\nu}=-\pd^i T_{i\nu}$, we can further simplify the expression as
\begin{align}
	\label{eq_B_op3}
	\mathcal{B}(t,\vx)=&\beta^\g (t,\vx)T_{0\g}(t,\vx)-\int_{-\infty}^{t} dt' \big[\pd^{\l}\beta^{\g}(t', \vx)T_{\l\g}( t',\vx)\no\\
	&-\pd^{i}\big(\beta^{\g}(t', \vx)T_{i\g}(t',\vx)\big)\big]
\end{align}

For the density operator $ \rho = Z^{-1} \exp [ -\int d^3 \vx \mathcal{B}(t,\vx) ] $, with $ Z = \text{Tr}\{\exp[...]\} $, the last term is just a surface integral that is typically assumed to be zero. Additionally, to make the ``covariance" structure of the theory more apparent, we specify the hypersurface $ d\Sigma^{\mu} = (dxdydz, \bm{0}) $ and use $ u^\mu = (1, \bm{0}) $ as the frame vector for the medium rest frame. With all this, the density operator can be expressed as:
\begin{align}
	\label{eq_zubco}
	\rho=&Z^{-1} \exp \Big[- [\int d\Sigma^{\l}\beta^\g (x)  T_{\l\g}(x)
	\no\\
	&-\int^{t}_{-\infty} d^4x'\pd^{\l}\b^{\g}(x')T_{\l\g}(x')]\Big]
\end{align}
The first term in the exponent represents the local equilibrium part, while the second term characterizes the non-equilibrium dynamical properties. Different choices for the frame vector and hypersurface~\cite{Liu:2021uhn, Becattini:2021suc} lead to different results. In this work, we do not explore these alternatives but instead select the ``almost" medium rest frame for the subsequent calculations, in which the flow vector is chosen as $u(x)=(1,\vu(x))$, where $\vu$ represents the infinitesimally small three components of the flow vector. We use $\vu$ to generate the gradients and then set it to zero. For simplicity, we also use the  same shorthand $u^\mu$ for $u^\mu(x)=(1,\vu(x))$ and $u^\mu = (1, \bm{0})$, where the distinction can be inferred from the context.

With this set up, we will expand  every thing into  linear order and re-express  Eq.~(\ref{eq_zubco}) in a more tractable form as
\begin{align}
	\label{eq_zub-mrest}
	&\rho=Z^{-1} \exp [ -\beta (H+ \tilde{H})]\;\;, \tilde{H}\equiv\tilde{H}_{\text{nd}}+\tilde{H}_{\text{d}}\no\\
	& \tilde{H}_{\text{nd}}=\frac{1}{\beta}\int d^{3}\vx \tilde{\beta}_\g(x) T^{0\g}(x)
	\no\\
	& \tilde{H}_{\text{d}}=-\frac{1}{\beta}\int^{t}_{-\infty} dt' \int d^{3}\vx'\pd_{\l}\beta_{\g}(x')T^{\l\g}(x')
\end{align}
with $\tilde{\beta}^\mu(x)=(\tilde{\beta}(x), \beta \textbf{u}(x))$, $\beta(x)=\beta+\tilde{\b}(x)$. $Z=\Tr\{\exp[-\beta (H+\tilde{H}_{\rm nd}+\tilde{H}_{\rm d})]\}$.
The $\beta$ is a constant inverse temperature $1/T$ and $\tilde{\b}(x)$ is a small term deviated from $\b$, and thus we have $\pd^\mu\tilde{\beta}^\nu=\pd^\mu\beta^\nu$ at leading order of gradients. We will directly apply the identity derived by Luttinger~\cite{Luttinger1964} (also see \cite{Hosoya:1983id}(a sign error in the paper)) to expand the exponential operators for small $ \tilde{H}_{\text{nd}}$ and $\tilde{H}_{\text{d}}$ into linear order terms
\begin{align}
	\label{eq_indentity}
&e^{-\beta (H+ \tilde{H})}=(1-\int^{\beta}_0 d\tau e^{-\tau H}(\tilde{H}_{\rm nd}+\tilde{H}_{\rm d})e^{H\tau}) e^{-\beta H}+...\no\\
&\rho=(1-\int^{\beta}_0 d\tau e^{-\tau H}\tilde{H}e^{H\tau}+\b\<\tilde{H}\>) \rho_G+o(\tilde{H})
\end{align}
in which the $\rho_G=e^{-\beta H}/Z_0$ with $Z_0=\Tr[e^{-\beta H}]$ and 
 $\langle..\rangle\equiv \text{Tr}\{...\rho_G\}=\text{Tr}\{...e^{-\beta H}\}/Z_0$. The last term in the bracket is the term comes from the denominator(trace), which will create cancellation between disconnected diagrams. 
The same formula can be obtained by using the Baker–Campbell–Hausdorff formula to expand the exponential in Eq.~(\ref{eq_zub-mrest}), when one carefully recombines the linear-order terms.
In this work, we focus on the linear order in gradients, with the study of higher-order effects deferred to future studies. 
Meanwhile, we will first calculate everything in the medium rest frame. Then, we will extrapolate the results to an arbitrary frame by assuming the covariance of the final expression, which is the standard procedure employed in the literature~\cite{Moore:2010bu, Hayata:2020sqz, Liu:2021uhn}.
\subsection{Non-dissipative part}
\label{sec_zuba_nd}
We will first study the non-dissipative part originating from $\tilde{H}_{\rm nd}$ in Eq.~(\ref{eq_indentity}) for a generic operator $O^a(t)$ with generic index $a$, where the deviation of its average value from global equilibrium is denoted as $\d O^{a}(t) \equiv \text{Tr}[O^a(\rho - \rho_G)]$.
There is a key identity  for later manipulation
\begin{align}
\label{eq_intdt}
&T^{0\g}(t,\vx) = \int^{t}_{-\infty} dt_o \pd_{t_o} T^{0\g}(t_o,\vx)+T^{0\g}(-\infty,\vx) \no\\
&= \int^{t}_{-\infty} dt_o \, i[H, T^{0\g}(t_o,\vx)]+T^{0\g}(-\infty,\vx)\,.
\end{align}
If we regulate $T^{0\m}(-\infty,\vx)=0$, we can obtain the non-dissipative part of $\d O^a$ as
\begin{align}
\label{eq_dO_nd1}
\delta O^{a}_{\rm nd}(t)=&\int d^{3}\vx \frac{\tilde{\beta}_\g(x)}{\beta}\langle -O^a(t)\int^{\beta}_0d\tau e^{- H\tau}T^{0\g}(x)e^{H\tau}\rangle\no\\
=&\int d^{3}\vx \int^{t}_{-\infty}dt_o\int^{\beta}_{0} d\tau\frac{\tilde{\beta}_\g(x)}{\b} \no\\
&\hspace{0.5cm}\times \langle O^a(t)e^{-H\tau}i[T^{0\g}(t_o,\vx),H]e^{H \tau}\rangle
\end{align}
With the same procedure in Eq.~(\ref{eq_dO_nd1}), it is not hard to show $\<\tilde{H}\>$ vanish with regulation $T^{0\g}(-\infty,\vx)=0$. Therefore, there is no trace correction and correction of the disconnected diagrams using this regulation. For those do not satisfy this regulation, in Sec.~\ref{sec_deco}, we provide an alternative derivation for Zubarev response approach with more considerations on this boundary term. In short, this term will not affect our calculations later.   

Then, in Eq.~(\ref{eq_dO_nd1}), we can integrate over $\tau$ first, rather than $t$, where the identity $\int^{\b}_0 d\t e^{- H\tau}[X,H]e^{ H\tau}=\int^{\beta}_{0}d\t\pd_{\t} (e^{- H\tau}X(t_o)e^{ H\tau })$ for a generic operator $X$ solves the integral. Meanwhile, we can obtain the relation $\langle O^a(t)e^{-\beta H}\tilde{H}_{\rm nd}(t)e^{\beta H }\rangle\equiv\langle\tilde{H}_{\rm nd}(t) O^a(t)\rangle$ by noticing the $e^{-\beta H}$ inside $\langle..\rangle$ and moving the $O^a$ around using the trace property. With all these steps, we can finally obtain a compact expression for $\delta O^a(t)$.
\begin{align}
	\label{eq_ndR}
	&\delta O^a_{\rm nd}(t)=\int d^{3}\vx \frac{\tilde{\beta}_\g(x)}{\beta}(-i)\langle O^a(t) \int^{t}_{-\infty}dt_o\big(T^{0\g}(t_o,\vx)\no\\
	&\qquad\qquad-e^{-\beta H}T^{0\g}(t_o,\vx)e^{\beta H }\big)\rangle\no\\
	&=\int d^{3}\vx \frac{\tilde{\b}_\g(x)}{\b}\int^{\infty}_{-\infty}dt_o(-i)\theta(t-t_o)\langle[O^a(t),T^{0\g}(t_o,\vx)]\rangle\no\\
	&\equiv\int d^{3}\vx \frac{\tilde{\b}_\g(x)}{\b}\int^{\infty}_{-\infty}dt_ou_\l G^{a\l\g}_R(t-t_o,\vx)
\end{align}
Following this convention,
 the retarded Green function for generic operators $X^{a}$ and $Y^{b}$ typically denoted  as $G^{ab}_{R} \equiv -i \theta(t - t_o) \langle [X^{a},Y^{b}] \rangle$.
In Eq.(\ref{eq_ndR}), we should note that the time variable in $\tilde{\b}(x)$ is $x = (t, \vx)$, which differs from the time variable in the operator $T^{0\g}(t_o, \vx)$. Additionally, as should be mentioned, Eq.(\ref{eq_ndR}) can also be derived from the typical linear response with adiabatic switching of the interactions~\cite{Kapusta:2006pm}. However, to obtain Eq.~(\ref{eq_ndR}), we need to treat the $t$ in $\tilde{\b}(t, \vx)$ as an external parameter, rather than as a dynamical time during the switching process.

This retarded Green function can then be expressed using the spectral functions as follows:
\begin{align}
	\label{eq_GR_iden}
&G_{R}^{a\l\g}(\o,\vx)=\int_{-\infty}^{\infty}d\o'\frac{1}{\pi}\frac{-\text{Im}G^{a\l\g}_R(\o',\vx)}{\o-\o'+i\e}\no\\
&G_{R}^{a\l\g}(t-t_o,\vx)=\int_{-\infty}^{\infty}\frac{d\o }{2\pi}e^{-i\o (t-t_o)}G_{R}^{a\l\g}(\o,\vx)
\no\\&=i \theta(t-t_o) \int^{\infty}_{-\infty}\frac{d\o'}{2\pi}e^{-i\o'(t-t_o)}2\text{Im}G^{a\l\g}_R(\o',\vx)
\end{align}
The last line uses residue theory to carry out the integral over $\omega$. Using the second line of Eq.(\ref{eq_GR_iden}) and Eq.(\ref{eq_ndR}) and integrating over $t_o$ to create a $\d(\o)$, the non-dissipative contribution $\delta O_{\rm nd}(t)$ becomes:
\begin{align}
	\label{eq_nd_final}
	\delta O^a_{\rm nd}(t)=\lim_{\o\rightarrow0}\int d^{3}\vx \frac{\tilde{\beta}_\g(t,\vx)}{\beta}u_{\l}G^{a\l\g}_R(\o,\vx) 
\end{align}
If both $O^a$ and  $ \tilde{H}_{\rm nd}$ are Hermitian operators, the limit of $G_R(\o,\vx)$ is just equal to the real part of it, i.e., $\lim_{\o\rightarrow0}\text{Re}G^{\l\nu}_R(\o,\vx)$ , and the imaginary part is zero. In time-reversal processes, although the retarded Green function $ G_R$ will become the advanced Green function $G_A$, their real parts are equal, $\text{Re}G_R = \text{Re}G_A$. Therefore, it is time-reversal even in this sense. As we should expect, the local equilibrium part of the density operator with $ \tilde{H}_{\rm nd}(t,\vx)$ has the time variable $t$, which is at the same time $t$ in the operator $ O^a(t)$ . Therefore, there is no time evolution between two operators, and it is equivalent to evaluating a static operator average in an ensemble, so that no dissipation linked to dynamics exists. The results in terms of the Green function are consistent with this intuitive expectation. In this sense, the time evolution in terms of $ t_o$ in the Green functions is auxiliary and just a trick for doing operator averages. This is also the reason we call it operator time $ t_o$ to distinguish it from the dynamical time $ t, t' $.

\subsection{Dissipative part}
\label{sec_zuba_d}
For the dissipative part, we insert the $\tilde{H}_d$ defined in Eq.~(\ref{eq_zub-mrest}) into Eq.~(\ref{eq_indentity}) (subtract the $\rho_G$). Then, following all the procedures in deriving the non-dissipative part, in the end, we just need to replace $\int^{t}_{-\infty}dt_o\int d^3\vx\tilde{\b}_{\g}(x) T^{0\g}(t_o,\vx)$ with $-\int^{t}_{-\infty} dt' \int d^{3}\vx'\pd_{\l}\beta_{\g}(x')\int^{t'} _{-\infty}dt_oT^{\l\g}(t_o,\vx')$  in the first and second line of Eq.~(\ref{eq_ndR}) and get the results as
\begin{align}
	\label{eq_od_time}
	\delta O^a_{\rm d}(t)=&-\frac{1}{\beta}\int^{t}_{-\infty} dt' \int d^{3}\vx' \pd_\l\beta_\g(x') \no\\
	&\times\int^{t'}_{-\infty}dt_oG^{a\l\g}_R(t-t_o,\vx')\,.
\end{align}
We then use $G^{a \l\g}_R=\langle O^a(t), T^{\l\g}(t_o,\vx')\rangle_{R}$ and its Fourier modes in Eq.~(\ref{eq_GR_iden}) to shorten the notation. In particular, with the last line in Eq.~(\ref{eq_GR_iden}), we obtain
\begin{align}
		\label{eq_od_o1}
	\delta O^{a}_{\rm d}(t)=&-\frac{1}{\beta}\int^{t}_{-\infty} dt' \int d^{3}\vx'\int^{\infty}_{-\infty} \frac{d\o'}{2\pi} \pd_\l\beta_\g(x') \no\\
	&\times \int^{t'}_{-\infty} dt_o e^{-i\o'(t-t_o)}(2i )\text{Im}G^{a\l\g}_R(\o',\vx')\no\\
	=&-\frac{1}{\beta}\int^{t}_{-\infty} dt' \int d^{3}\vx' \pd_\l\beta_\g(x')\no\\
	&\times \int^{\infty}_{-\infty}\frac{ d\o' }{2\pi} e^{-i\o'(t-t')}\frac{2\text{Im}G^{a\l\g}_R(\o',\vx')}{\o'}\,,
\end{align}
where the boundary term at infinity is regulated to zero. 
We then take slow (zero frequency) limits, and keep only the leading term $\lim_{\o'\rightarrow0}2\pd_{\o'}\text{Im}G^{\mu\nu}_R(\o',\vx')$, while neglecting all other $\o'$ dependence (the higher-order terms in $\o'$). The whole $\o'$ dependence will come from $e^{-i\o'(t-t')}$, and integrating over $\o'$ will generate a delta function in time as
\begin{align}
	\label{eq_od-delta}
	\delta O^a_{\rm d}(t)=&-\frac{1}{\beta}\int^{\infty}_{-\infty} dt'\theta(t-t') \int d^{3}\vx' \pd_\l\beta_\g(x')\delta(t-t') \no\\
	&\times\lim_{\o'\rightarrow0}2\frac{\pd}{\pd \o'}\text{Im}G^{a\l\g}_R(\o',\vx')
\end{align}
It should be noted the $\int_{-\infty}^{\infty}dt'\theta(t-t')\delta(t-t')=\int d\o (i/(2\pi))(\o+i\e)^{-1} =1/2$ in Fourier space with $\e$ regulation. This cancels the $2$ in the second line of Eq.~(\ref{eq_od-delta}), and the  Eq.~(\ref{eq_od-delta}) becomes the familiar linear response formula:
\begin{align}
		\label{eq_od-final}
	&\delta O^a_{\rm d}(t)=\no\\
	&-\frac{1}{\beta} \int d^{3}\vx' \pd_\l\beta_\g(t,\vx') \lim_{\o'\rightarrow0}\frac{\pd}{\pd\o'}[\text{Im}G^{a\l\g}_R(\o',\vx')]
\end{align}
As one should notice, the entire term depends on the imaginary part of the retarded Green's function. In time-reversal processes, the retarded Green's function $G_R$ becomes the advanced Green's function $G_A$, and their imaginary parts satisfy the relation $\text{Im}G_R = -\text{Im}G_A$, which generates a minus sign and makes the coefficients T-odd. This is also consistent with expectations from dissipative contributions.

\subsection{Analytical continuation,  Euclidean time and lattice QFT}
\label{sec_conti}
The discussion in the previous section connects the Zubarev formalism to the retarded Green functions. As the correlation function only depends on two time variables, with translation symmetry in the time direction, its analytical structure is quite simple and can be related to, or more exactly, analytically continued to Euclidean/imaginary time space. Then, we can either use the diagrammatic technique to perform perturbation theory or employ non-perturbative methods such as the Dyson-Schwinger equation~\cite{Roberts:2000aa}, functional renormalization group~\cite{Fu:2019hdw}, or lattice QCD to evaluate the correlation function in Euclidean time at first. With the results in Euclidean space, we perform the analytical continuation back to real time for the final results.

Using the Kallen-Lehmann representation and splitting the diagonal and off-diagonal parts, we have the matrix elements of operators as $\<m|X|n\>=X_{mn}+X_m\delta_{mn}$, $\<m| Y|n\>=Y_{mn}+Y_m\delta_{mn}$ for two bosonic operators. 
For the time-dependent operator $X(t)=e^{iHt}Xe^{-iHt}$, we can express the correlation functions as
\begin{align}
	\label{eq_GLSK}
	&\<X(t)Y(0)\>=Z^{-1}_0e^{-\b E_n}X_{nm}Y_{mn}e^{-i(E_m-E_n)t}+\mathcal{C} \no\\
	&\<Y(0)X(t)\>=Z^{-1}_0e^{-\b E_m}Y_{mn}X_{nm}e^{-i(E_m-E_n)t}+\mathcal{C} \no\\
	&\mathcal{C}=\sum_m[\rho_G]_m X_mY_m\rightarrow\mathcal{C}=\<X\>\<Y\> \;\text{some cases}\no\\
	&\<[X(t),Y(0)]\>=\int d\o A(\o) e^{-i\o t}
\end{align}
where we will briefly explain ``some cases" in Sec.~\ref{sec_deco}.  The spectral function in Eq.~(\ref{eq_GLSK}) is defined to be
\begin{align}
	\label{eq_spe_gen1}
		A(\o)=&Z_0^{-1}\sum_{m,n}(e^{-\b E_n}-e^{-\b E_m})X_{nm}Y_{mn}\no\\
		&\times\delta(\o-(E_m-E_n))
\end{align}
For retarded Green function, it will be
\begin{align}
	\label{eq_GR_leh}
	&G_R(t)=-i\theta(t)\< [X(t),Y(0)]\>\equiv-i\theta(t)\int d\o A(\o) e^{-i\o t}\no\\
	&G_R(\o)\equiv\int dt e^{i\o t} G(t)=\int d\o' \frac{A(\o')}{\o-\o'+i\e}
\end{align}
With Eq.~(\ref{eq_GR_leh}), according to our convention the imaginary part of $G_R$ and the spectral functions are related as
\begin{align}
	\label{eq_spe_gen2}
	&A(\o)=\frac{-\text{Im }G_R(\o)}{\pi}
\end{align}
For imaginary time operators and correlation functions, we have $X(\tau)=e^{H\t }X e^{-H\tau}$   and $G(\t)=-\< T_{\t}[X(\tau)Y(0)]\>=-\< X(\tau)Y(0)\>$ within $\tau\in[0,\b]$, which can be represented in spectral representation as
\begin{align}
	\label{eq_Gim_leh}
G(\t)&=-Z_0^{-1}e^{-\b E_n}X_{nm}Y_{mn}e^{-(E_m-E_n)\t}-\mathcal{C}\no\\
G(i \o_n)&=\int^{\b}_0 d\t  e^{i\o_n \t}G(\t)\no\\
&=Z^{-1}_0\frac{(e^{-\b E_l}-e^{-\b E_m})(X_{lm}Y_{ml})}{i\o_n-(E_m-E_l)}-\b\mathcal{C}\d_{n0}\no\\
&\equiv\int d \o' \frac{A(\o')}{i\o_n-\o'}-\b\mathcal{C}\d_{n0}
\end{align}
where the $i\o_n=2 n\pi/\beta$ (bosonic) is the Matsubara frequency.
Comparing the last line of Eq.~(\ref{eq_Gim_leh}) and Eq.~(\ref{eq_GR_leh}), we can identify that analytical part of $G_R(\o)$ and $G(i\o_n)$ will have the relation 
\begin{align}
	\label{eq_conti_final}
&G_R(\o)=G(i \o_n)|_{i\o_n\rightarrow \o+i\e}
\end{align}
with non-analytical part $\mathcal{C}$ neglected.

This formula in Eq.~(\ref{eq_conti_final}) is mostly used for the case that the analytical formula is known. 
For numerical continuation, the results are discussed later. Making it more specific for the $T^{\mu\nu}$ and $O^{a}$ operators, we define the Euclidean Green function $G^{\mu\nu}$ as
\begin{align}
		G^{a\mu\nu}(\tau,\vx )=(-)\langle T_\t [O^a(\tau) T^{\mu\nu}(0,\vx)]\rangle\,. 
\end{align}
With this Euclidean-time Green function, the non-dissipative/local equilibrium part of the contribution, i.e., Eq.~(\ref{eq_ndR}), becomes
\begin{align}
	\label{eq_nd_E}
	\delta O^a_{\rm nd}(t)=&\int d^{3}\vx \frac{\tilde{\b}_\g(x)}{\b}\int^{\beta}_{0}d\tau\, u_{\l}\big[ G^{a\l\g}(\tau,\vx )\no\\
	&+\<O^a\>\<T^{\l\g}(\vx)\>\big]
\end{align}
The last term is originated from the trace correction at Eq.~(\ref{eq_indentity}).
As shown, for the non-dissipative contribution, everything is well-defined in Euclidean space. As already demonstrated in many non-perturbative approaches, such as the DSE , FRG , and lattice QFT, calculations in Euclidean space are much simpler and more stable compared to real-time calculations. This gives us a good chance to study the related response using non-perturbative methods. Especially, with lattice QCD, a wide range of observables can be calculated from first principles.

For the dissipative contribution, a direct real-time calculation or analytical continuation is unavoidable since it involves the true dynamics in non-equilibrium states. It is well known that the Euclidean-time Green function can be connected to the real-time/frequency spectral function as (see Ref.~\cite{Liu:2017qah} and references therein):
\begin{align}
&G(\tau)=\frac{1}{\beta}\sum_n e^{-i\o_n\tau}G^{\mu\nu}(i\o_n)\no\\
&=-\int^{\infty}_{-\infty} d\o \frac{e^{\beta \o }e^{-\o\tau}}{e^{\beta \o}-1}\frac{-1}{\pi}\text{Im}G_R(\o)\no\\
&\equiv-\int^{\infty}_{0} d\o \frac{\cosh[(\tau-\beta/2) \o]}{\sinh[\beta\o/2]}(\frac{-1}{\pi})\text{Im}G_R(\o)
\end{align}
Mathematically, it is an integral equation that can be solved by taking the inverse of the kernel $K(\o,\tau)=\cosh[(\tau-\beta/2)\o]/\sinh[(\beta/2)\o]$, reconstructing $\text{Im}G^{\mu\nu}_R(\o)=\pi \int d \t K^{-1}(\o,\tau)G(\tau)$. However, in practice, the properties of $K^{-1}$ are highly problematic. In the diagonal basis, the kernel contains eigenvalues close to zero~\cite{Shi:2022yqw}, which means its inverse will be close to infinite. This simply implies that small errors in some specific modes will be infinitely enlarged when using $K^{-1}$ to map $G(\tau)$ to reconstruct $\text{Im}G_R(\o)$. This basically states that the inverse/analytical continuation back to real time is fundamentally difficult. Nevertheless, with some ansatz, we can still extract useful information from this reconstruction. For this approach, we refer to a most recent attempt for such a procedure for the heavy-quark diffusion coefficient~\cite{Altenkort:2023eav}, where one can gain insight into how these methods work by reading the paper and the references cited therein.
\subsection{Alternative derivation to Zubarev response approach}
\label{sec_deco}
As shown in Eq.~(\ref{eq_indentity}) and Eq.~(\ref{eq_dO_nd1}), the key to proceed is to handle the correlation function with both imaginary time and real time. If we define that $G(\t-\t',t-t')=-\<T_\t [X(\t,t)Y(\t',t')]\>$ where $X(\t,t)=e^{H\t}X(t)e^{-H\t}$, following the same procedure in calculating Eq.~(\ref{eq_Gim_leh}), we can derive a compact formula for $G(\t-\t',t-t')$ as 
\begin{align}
	\label{eq_GEt}
	&G(\t-\t',t-t')\no\\&=\frac{1}{\b}\sum_{n}\int \frac{d\o}{2\pi}  e^{-i\o_n (\t-\t')} e^{-i\o(t-t')}G(i\o_n,\o)\no\\
	&G(i\o_n,\o)=2\pi[\frac{A(\o)}{i\o_n-\o}-\b\mathcal{C}\delta(\o)\d_{n0}]
\end{align}
The spectral function $A(\o)$ above is defined in Eq.~(\ref{eq_spe_gen1}) and  Eq.~(\ref{eq_spe_gen2}). 
With the relations in Eq.~(\ref{eq_GEt}) and noting that $-\<T_\t [X(0,t)Y(-\t,t')]\>=-\< X(0,t)Y(-\t,t')\>$, we can calculate the integral $\int^{\b}_0 d\t$ of the $G(\t-\t',t-t')$, which is of the same structure as the integral in the first line of Eq.~(\ref{eq_dO_nd1}), in the following form:
\begin{align}
	\label{eq_intbeta}
	&-\int^{\b}_0d\t\< X(0,t)Y(-\t,t')\>=\int^{\b}_0d\t G(\t,t-t')\no\\
&= \int d\o\frac{A(\o)}{-\o}e^{-i\o(t-t')}-\b\mathcal{C}\no\\
&\equiv\int \frac{d\o}{2\pi}\frac{2\text{Im}G_R(\o)}{\o}e^{-i\o(t-t')}-\b\mathcal{C}
\end{align}
Inserting $\tilde{H}_{\text{nd}}$ from Eq.~(\ref{eq_zub-mrest}) and $O^{a}$ into Eq.~(\ref{eq_intbeta}), we can obtain
\begin{align}
	\label{eq_nd_final2}
	\delta O^a_{\rm nd}(t)&=\int d^{3}\vx \frac{\tilde{\beta}_\g(t,\vx)}{\beta}u_{\l}\int d\o \frac{1}{-\o}\frac{-\text{Im}G^{a\l\g}_R(\o,\vx) }{\pi}\no\\ 
	&\equiv\lim_{\o\rightarrow0}\int d^{3}\vx \frac{\tilde{\beta}_\g(t,\vx)}{\beta}u_{\l}G^{a\l\g}_R(\o,\vx) 
\end{align}
The second line is obtained with the help of Eq.~(\ref{eq_GR_leh}) and Eq.~(\ref{eq_spe_gen2}). The static correlation term $-\mathcal{C}+\<O^a\>\<\tilde{H}_{\rm nd}\>$ in the second line is neglected temporarily for the last line, and we will discuss how it contributes in the end.
Using Eq.~(\ref{eq_intbeta}), Eq.~(\ref{eq_zub-mrest}) and Eq.~(\ref{eq_indentity}), and  following a similar procedure, the dissipative part can be expressed as
\begin{align}
	\label{eq_od_final2}
	&\delta O^{a}_{\rm d}(t)=-\frac{1}{\beta}\int^{t}_{-\infty} dt' \int d^{3}\vx' \pd_\l\beta_\g(x')\\
	&\qquad\qquad\int^{\infty}_{-\infty}\frac{ d\o' }{2\pi} e^{-i\o'(t-t')} \frac{2\text{Im}G^{a\l\g}_R(\o',\vx')}{\o'}\no\\
	&=-\frac{1}{\beta} \int d^{3}\vx' \pd_\l\beta_\g(t,\vx') \lim_{\o'\rightarrow0}\frac{\pd}{\pd\o'}[\text{Im}G^{a\l\g}_R(\o',\vx')]\no
\end{align}

The only remaining term to be evaluated is the static correlation term that is typically neglected or vanish. To proceed, we will use an alternative expression for Eq.~(\ref{eq_B_op1}) rather than Eq.~(\ref{eq_B_op3}), which is
\begin{align}
\tilde{H}=\frac{1}{t-t_0}\int^{t}_{t_0} dt' \frac{1}{\beta}\int d^{3}\vx' \tilde{\beta}_\g(x') T^{0\g}(x')
\end{align}
where we switch from Abel regularization $\e\int^{t}_{-\infty}dt'e^{-\e(t-t')}$~\cite{zubarev1974nonequilibrium} to a more intuitive time-average regularization (note $\tilde{\b}(x)=\b(x)-\b$). 

Then, we denote the static correlation of the $T^{\m\n}$ and $O^a$ as 
\begin{align}
	\label{eq_Gconst}
&G_{\mathcal{C}}^{a\l\g}(\vx')=\b\<O^{a}\>\<T^{\l\g}(\vx')\>-\b\mathcal{C}^{a\l\g}(\vx')\\
	\label{eq_Const}
&\mathcal{C}^{a\l\g}(\vx')=\sum_m [\rho_G]_m [O^{a}]_m [T^{\l\g}(\vx')]_m\,.
\end{align}
With this static correlation, we will have an additional contribution to the operator average $\d O^{a}_{\rm cs}(t)$, which is
\begin{align}
	 \label{eq_cs_sp}
 &\d O^{a}_{\rm cs}(t)=\frac{1}{t-t_0}\int^{t}_{t_0}dt' d^3\vx' \frac{\tilde{\b}_\g(t',\vx')}{\b}u_{\l}G_{\mathcal{C}}^{a\l\g}(\vx')\no\\
 &=\int d^3\vx'\frac{\tilde{\beta}_\g(t,\vx')}{\b}u_{\l}G_{\mathcal{C}}^{a\l\g}(\vx')- \frac{1}{t-t_0}
\no \\
 &\times\int^{t}_{t_0}dt' \int d^3\vx'(t'-t_0)\frac{\pd_0{\tilde{\b}_\g(t',\vx')}}{\b}u_{\l}G_{\mathcal{C}}^{a\l\g}(\vx')\\
   \label{eq_cs_av}
 &=\int d^3\vx'\frac{\overline{[\tilde{\beta}]}_\g(t,\vx')}{\b}u_{\l}G_{\mathcal{C}}^{a\l\g}(\vx')\\ &\overline{[\tilde{\beta}]}_\g(t,\vx')\equiv\frac{1}{t-t_0}\int^{t}_{t_0}dt' \tilde{\beta}_\g(t',\vx')\no
\end{align}
The last term indicates that it has memory effects, which is not surprising that since the Zubarev's formalism can construct transport equations including memory effects. However, for the examples discussed in this paper, we have a field theory that has $[O^a]_m=\<O^{a}\>$,$[T^{\m\n}]_m=\<T^{\m\n}\>$ and Eq.~(\ref{eq_Gconst}) turns out to be zero, which is a manifestation of the cancellation between the disconnected diagrams and it is a de-correlations at infinitely separated time\cite{Hosoya:1983id}.
If there is nonzero $G^{a\l\g}$, which means that $\mathcal{C}^{a\l\g}$ is correlated static trace,  then this correlation function $\mathcal{C}^{a\l\g}$ can be calculated as (Eq.~(\ref{eq_Gim_leh}) and Eq.~(\ref{eq_GR_leh}))
\begin{align}
\mathcal{C}^{a\l\g}=G^{a\l\g}(i\o_n=0)-\lim_{\o\rightarrow0}G^{a\l\g}_R(\o)
\end{align}
This formula will subtly rely on the spectral properties and real-time properties. Therefore, extracting $\mathcal{C}^{a\l\g}$ will be challenging on lattice QFT. On the other hand, we can still separate it into two contributions in Eq.~(\ref{eq_cs_sp}) and interpret the first term as a non-dissipative term. Thus, we do not need to separate two zero modes. Therefore, the non-dissipative term becomes those shown in Eq.~(\ref{eq_nd_E}), which can be directly evaluated on lattice QFT or other imaginary time Monte-Carlo methods. 

\section{Vector polarization for spin-1/2}
Starting from this section, we will apply the Zubarev response approach discussed in the previous section to polarization phenomena in local equilibrium. The first example, as discussed in this section, is the vector polarization for spin-1/2 particles.

The polarization for spin-1/2 particles is related to the famous $\Lambda$ hyperon polarization~\cite{STAR:2017ckg}, especially its puzzling local polarization~\cite{Adam:2019srw}. One major progress in understanding this puzzle is the discovery of shear-induced polarization~\cite{Liu:2021uhn,Fu:2021pok,Becattini:2021iol,Becattini:2021suc}. The two theoretical papers, one following the Zubarev approach using operator commutation techniques~\cite{Becattini:2021suc}, while the other~\cite{Liu:2021uhn} is based on Luttinger's perspective of linear response~\cite{Luttinger1964} with diagrammatic techniques and using a metric dictionary to connect the shear with the metric( gradients). In the following, we will repeat the calculations in the Zubarev response approach but using diagrammatic techniques rather than operator commutation.
\label{sec_spinhalf}
\subsection{Main formalism}
We begin the derivation by presenting the energy-momentum tensor~\cite{Liu:2021uhn} of the spin-1/2 free fermion as follows:
\begin{align}
	\label{eq_Tmn_half}
	T^{\l\g}=\frac{i}{4}\bar{\psi}(\gamma^\l\overset{\leftrightarrow}{\partial^\g}+\gamma^\g\overset{\leftrightarrow}{\partial^\l})\psi
\end{align}
where $ f_2 \overset{\leftrightarrow}{\pd_{\mu}} f_1 =f_2(\pd_{\mu} f_1)-(\pd_{\mu} f_2)f_1$.
This is the Belinfante form of the energy-momentum tensor of the spin-1/2 fermion, and we will not explore the problem of the pseudo-gauge~\cite{Becattini:2018duy} in this work. 
The operator we will calculate is $J^{\m}_5$, which is the Pauli-Lubanski definition of the spin vector times mass.
Its Wigner form  $\hat{\mathcal{J}}^\mu_5(t,\vx,\vp)$ can be expressed as:
\begin{align}
	\label{eq_J5de}
	&\hat{\mathcal{J}}_5^\mu(t,\vx,\vp)=\ve_{\vp}\int d^3\vy e^{-i\vp\cdot\vy}\bar{\psi}\left(t,\vx_{-}\right)\gamma^\mu\gamma^5\psi\left(t,\vx_{+}\right) 
\end{align}
where $\vx_{\pm}=\vx\pm\vy/2$. 
For  Eq.~(\ref{eq_nd_final}), we replace the generic operator $O^a$ with the $\mathcal{J}_5^{\mu}(t,\vx,\vp)$ and the $G_R$ inside with $ G_R^{\mu\l\g}(\o,\vp,\vq)$ defined as follows
	\begin{align}
		\label{eq_Gdefpq}
		&G_R^{\mu\l\g}(t-t_o,\vx,\vx',\vy)\equiv\int \frac{d\o}{2\pi}\int \frac{d^3\vq }{(2\pi)^3}\frac{d^3\vp }{(2\pi)^3}e^{-i\o(t-t_o)}\,\no\\
		&\hspace{3.5cm}\times e^{i\vp\cdot \vy}e^{i\vq\cdot (\vx-\vx')} G_R^{\mu\l\g}(\o,\vp,\vq)\,\no\\
		&\equiv -i\theta(t-t_o)\big\langle [\bar{\psi}\left(t,\vx_{-}\right)\gamma^\mu\gamma^5\psi\left(t,\vx_{+}\right)
		,T^{\l\g}(t_o,\vx')]\big\rangle
	\end{align}
With these replacements, we can re-express the Eq.~(\ref{eq_nd_final}) for the $\mathcal{J}_5^{\mu}(t,\vx,\vp)$  as
\begin{align}
	\label{eq_j5-half-gen-0}
	&\mathcal{J}_5^\mu(t,\vx,\vp)\\
	&=\lim_{\o\rightarrow0}\frac{\ve_{\vp}}{\beta}\int d^3\vx'\frac{ d^3\vq}{(2\pi)^3} G_R^{\mu0\gamma}(\o,\vp,\vq) e^{i\vq\cdot (\vx-\vx')}\tilde{\b}_\g(t,\vx')\no\\
	&=\lim^{\vq\rightarrow0}_{\o\rightarrow0}\frac{-i\ve_{\vp}}{\beta}\frac{\pd}{\pd \vq^{i}} G_R^{\mu0\g}(\o,\vp,\vq)\pd_i \tilde{\b}_\g(x)+o(\pd)\no
\end{align}
To obtain the second line, we expand $G_R$ in terms of the power series of $\vq$ and only take the linear order of $\vq$ (as the zeroth order turns out to vanish in the end). Therefore, the entire $\vq$ dependence comes from $\vq^i\tilde{\b}(x') e^{i \vq\cdot(\vx-\vx')}$. Integrating this over $\vq$ with partial integration once, we obtain a term $-i \pd_i \tilde{\b}_\g\delta(\vx-\vx')$. From here, we can see that neglecting higher powers of $\vq$ is equivalent to neglecting higher orders of derivatives $o(\pd)$. 

For the notation of the limit ``$\lim$", we take the lower limit first and then the upper limit, i.e., the slow limit in Ref.~\cite{Liu:2021uhn}. With the frame vector $u^\mu$ in the medium rest frame and the projector $\bar{\Delta}^{\m\n}=\eta^{\m\n}-u^\mu u^\n$, we can define the transverse or longitudinal of objects as $p^\mu_{\perp}=(0,\vp)=\Delta^\mu_{\nu}p^\nu$ and $\pd_{\m}^{\perp}=(0, \bm{\pd})=\Delta^{\nu}_{\mu}\pd_\nu$, $\ve_{u}= p\cdot u=\ve_{\vp}$  With these notations, we can make Eq.~(\ref{eq_j5-half-gen-0}) look more ``covariant" as
\begin{align}
	\label{eq_j5-half-gen}
	\mathcal{J}_5^\mu(t,\vx,\vp)
	=\lim^{\vq\rightarrow0}_{\o\rightarrow0}\frac{i\ve_{u}}{\beta}\frac{\pd}{\pd q^{\perp}_\xi}u_\l G_R^{\mu\l\g}(\o,p_\perp,q_\perp)\pd_\xi \tilde{\b}_\g(x)
\end{align}
From here on, we will neglect $o(\pd)\no$, with the understanding that all calculations are linear order in gradients.

The Euclidean Green function  $G^{\mu\l\g}(\tau-\tau^{\prime},\vx,\vx',\vy)$, which shares the same spectral function with the $G^{\mu\l\g}_R$  discussed above, can be expressed as
\begin{align}
	\label{eq_j5-half-gen-1}
	 &G^{\mu\l\g}(\tau-\tau^{\prime},\vx,\vx',\vy)\no\\
	&= -\left\langle T_{\tau}\left[\bar{\psi}\left(\tau,\vx_{-}\right)\gamma^\mu\gamma^5\psi\left(\tau,\vx_{+}\right)T^{\l\gamma}(\tau',\vx')\right]\right\rangle\no\\
	&=-\big\langle T_\tau[ \bar{\psi}\left(\tau,x_{-}\right)\gamma^\mu\gamma^5\psi\left(\tau,\vx_{+}\right)\no\\
	&\qquad\frac{i}{4}\bar{\psi}\left(\tau^{\prime},\vx'\right)\left(\gamma^{\l}\overset{\leftrightarrow}{\pd^{\prime\g}}+\gamma^{\g}\overset{\leftrightarrow}{\pd^{\prime\l}}\right)\psi\left(\tau^{\prime},\vx'\right)]\big\rangle
\end{align}
where the $T_\t$ denotes the imaginary time-ordering and $T^{\l\g}$ in Eq.~(\ref{eq_Tmn_half}) is employed and $ \partial_t =i\partial_\tau $.

The correlation function has non-perturbative meaning, even in terms of the diagrammatic language, as will be further elaborated in Sec.~\ref{sec_skeleton}. However, here, we only consider the leading order of the skeleton expansion (see Sec.~\ref{sec_skeleton}), which is
\begin{align}
	\label{eq_j5_SS}
	&G^{\m\l\g}(\tau-\tau^\prime,\vx,\vx',\vy)=
	\frac{i}{4}\text{Tr}[\gamma^\mu\g^5S(\tau-\tau^\prime,\vx+\frac{\vy}{2}-\vx')\no\\
	&\times\left(\gamma^{\l}\overset{\leftrightarrow}{\tilde{\pd}^{\prime\g}}+\gamma^{\g}\overset{\leftrightarrow}{\tilde{\pd}^{\prime\l}}\right)S(\tau^\prime-\tau,\vx'-\vx+\frac{\vy}{2})],
\end{align}
where the propagators $S$ are the full dressed propagators in the skeleton expansion. However, to recover our previous results in Ref.~\cite{Liu:2021uhn}, we enforce the full propagator to be the simplest ``free" and bare one in this section, which is
\begin{align}
	\label{eq_S_def}
	S(\tau,\vx)\equiv&\frac{1}{\beta}\sum_{\o_n}\int \frac{d^3\vp}{(2\pi)^3}\int d\o \,s_\o(p_k\g^k+m )\no\\
	&\times\delta(\o^2-\varepsilon^2_{\vp})\frac{1}{i\o_n-\o}e^{-i\omega_n\tau +i\vp\cdot\vx}.
\end{align}
where the $ s_\o $ is the sign of $ \omega $. The $\o_n=(2n+1)\pi/\b$ represents the Matsubara frequency for fermions. For the time derivatives,  the $ \partial_t =i\partial_\tau $ which will generate an $-i(i\o_n)$ on the $e^{-i\omega_{n}\tau+i\boldsymbol{p}\cdot\boldsymbol{x}}$. An \textbf{important subtlety }to note is that the time derivative of correlation function will generate a $\delta(\tau)$ due to the time-ordered products in the $S$. However, according to the last line of Eq.~(\ref{eq_j5-half-gen-1}), the time derivatives do NOT act on these $\theta$ functions, and this artificial $\delta(\tau)$ should be removed. 
In general, $\tilde{\pd}$ represents derivatives with time derivatives ``on shell", which requires using the equation of motion for $S^>$ to replace the time derivatives with spatial derivatives and self-energies, thereby removing the $\delta(\tau)$. 
However, if we use $ \tilde{\partial}_t$ to generate an $-i\o$ instead of $-i(i\o_n)$, we will find terms like $\sum_n e^{-i\o_n \tau }i\o_n/(i\o_n-\o)=\sum_n e^{-i\o_n \tau }\o/(i\o_n-\o)+\b\delta(\t)$. Replacing $i\o_n$ with $\o$ is equivalent to removing the $\d(\t)$ mentioned earlier. Therefore, the on-shell derivative with $\d(\t)$ removed is
\begin{align}
	\label{eq_S_deriv}
	\tilde{\pd}^{\prime\l}S(\tau,\vx)\equiv&\frac{1}{\beta}\sum_{\o_n}\int \frac{d^3\vp}{(2\pi)^3}\int d\o \,s_\o(-i p^{\l})(p_k\g^k+m )\no\\
	&\times\delta(\o^2-\varepsilon^2_{\vp})\frac{1}{i\o_n-\o}e^{-i\omega_n\tau +i\vp\cdot\vx}
\end{align}
 where momentum $p$ is the more familiar real-time four-momentum $p=(\o,\vp)$ inside the integral. With all these notations, we will express the details of the calculation of Eq.~(\ref{eq_j5_SS}) as below:

	\begin{align}
		\label{eq_half_longG}
		&G_R^{\mu\l\gamma}(\tau-\tau^\prime,\vx,\vy,\vx')= \frac{i}{4}
		\frac{1}{\beta}\sum_{\omega_n}\frac{1}{\b}\sum_{\nu_n}\int\frac{d^4 k}{(2\pi)^3}\frac{d^4 k^\prime}{(2\pi)^3}
		\no\\&e^{-i(\o_n+\nu_n)(\tau-\tau^\prime)+i\vk\cdot(\vx+\frac{\vy}{2}-\vx')} e^{-i\nu_n(\tau^\prime-\tau)+i\vk^\prime\cdot(-\vx+\frac{\vy}{2}+\vx')}\no\\
		& \times s_{k_0}s_{k_0'}\frac{\delta(k_0^2-\ve_{\vk}^{2})}{i\o_n+i\nu_n-k_0}\frac{\delta(k_0^{\prime2}-\ve_{\vk^\prime}^{2})}{i\nu_n-k_0'}\text{Tr}\Big[\g^\mu\g^5(k_\rho\g^\rho+m)
		\no\\
		&[\gamma^{\l}(-ik-ik^\prime)^\g+\gamma^{\g}(-ik-ik^\prime)^{\l}](k^{\prime}_{\sigma}\g^\s
		+m)\Big]
	\end{align}

To evaluate the traces of the $\g$ matrices in the last line of Eq.~(\ref{eq_half_longG}), we will need to employ the trace identity (with the convention $\ve^{0123}=1$) as
\begin{align}
	\label{eq_gamma_tr}
	\text{Tr}\left[\g^\mu\g^5(k_\rho\g^\rho+m)\g^{\xi}(k^{\prime}_{\s}\g^\s+m)\right]=k_\r k_\s^{\prime}(-4i\ve^{\r\xi\s\mu})\,.
\end{align}
The $\xi$ in Eq.~(\ref{eq_gamma_tr}) will be replace by $\gamma$ or $\l$ when applied to Eq.~(\ref{eq_half_longG}).

In addition to the evaluation of the trace, we also need to perform the summation over $\nu_n$ using well-known contour integration (see Refs.~\cite{Bellac:2011kqa,fetter2012quantum}), which essentially is
\begin{align}
	\label{eq_contour_int}
	&\frac{1}{\b} \sum_{\n_n}\frac{1}{i\o_n+i\n_n-k_0}\frac{1}{i\n_n-k'_0}\no\\
	&=\mp\frac{1}{2\pi i}\oint n(z) \frac{1}{i\o_n+z-k_0}\frac{1}{z-k'_0}\no\\
	&=\pm \frac{n(k'_0)-n(k_0)}{i\o_n+k'_0-k_0}
\end{align}
The final contour is clockwise when going through the circle at infinite radius. This generates an additional minus sign when moving from the second to the third line. Meanwhile, the upper/lower sign corresponds to fermionic ($(2n+1)\pi/\b$) or bosonic ($2n\pi/\b$) for $\nu_n$. After performing this summation, we are allowed to analytically continue $i\o_n$ to $\tilde{\o}+i\e$ for the retarded Green function. We will use the same notation $n$ for Bose-Einstein, Fermi-Dirac, and Boltzmann distributions. It is straightforward to identify which is exactly used along with the context; otherwise, we will provide clarification.

Then, we further perform the integral over $k_0$ and $k'_0$, but only for positive frequency with $s_{k_0} = +1$ and $s_{k'_0} = +1$, rather than taking all four combinations of $s_{k_0}$ and $s_{k'_0}$. The reason for only selecting this mode is that the relevant physics of polarization is contributed from the particle density-density correlation, and discussions for mode selection can be found in Sec.~\ref{sec_mode}. With $q = k - k'$ and $p = (k + k')/2$, so that $k | k' = (\ve_{\vp \pm \vq/2}, \vp \pm \vq/2)$ ($``|" = \text{OR}$), we can take the Fourier-transformed Green function $G_R^{\mu\l\gamma}(\tilde{\o}, \vp, \vq)$ out of Eq.~(\ref{eq_half_longG}) following the definitions in Eq.~(\ref{eq_Gdefpq}), which is
\begin{align}
G_R^{\mu\l\gamma}(\tilde{\o},\vp,\vq)&=	
	\frac{-2i}{4\ve_{\vk}\ve_{\vk^\prime}} \frac{n(\ve_{\vk^\prime})-n(\ve_{\vk})}{\tilde{\o}+\ve_{\vk^\prime}-\ve_{\vk}+i\e} \no\\
	&\times\left[k_\r k^{\prime}_\s\ve^{\r\g\s\mu}p^{\l}+k_\r k^{\prime}_\s\ve^{\r\l\s\mu}p^\g \right]\,.
\end{align}
Then, we set $\o = 0$ and expand the expression to leading order in $q^\mu = (\vp \cdot \vq / {\varepsilon_{\vp}}, \vq)$. After performing these steps and neglecting all higher orders in $\vq$, we arrive at a simplified expression as follows:
\begin{align}
	&G_R^{\mu\l\gamma}(0,\vq,\vp)=
\frac{i}{\ve_{u}^2}
	\frac{\pd n(\ve_{u})}{\pd\ve_{u}} p_\r q_\s\ve^{\r\s\mu(\g}p^{\l)}\no\\
	&=\frac{i}{\ve_{u}^2}\frac{\pd n(\ve_{u})}{\pd\ve_{u}} p_\r (q^{\perp}_\sigma+(q\cdot u)u_{\s})\ve^{\r\s\mu(\g}p^{\l)}
\end{align}
To evaluate the derivatives of $q_\perp$, we split $q^{\s}$ into the longitudinal part $(q \cdot u) u^{\s}$ and the transverse part $q_\perp^{\s}=(0,\vq)$, following similar logic discussed around Eq.~(\ref{eq_j5-half-gen}), where $q\cdot  u=-p_\perp\cdot q_\perp/\ve_u=\vp\cdot\vq/\ve_{\vp}$. 
Meanwhile, we also have the identity of the differential on $q^\s$ as
\begin{align}
	\label{eq_dperp}
\frac{\pd}{\pd q^\perp_{\xi}}(q^{\perp}_\sigma+(q\cdot u)u_{\s})=\bar{\Delta}^{\xi}_{\sigma}-\frac{p_\perp^{\xi}}{\ve_{u}}u_{\s}
\end{align}
Then, using  Eq.~(\ref{eq_j5-half-gen}) and Eq.~(\ref{eq_dperp}), we obtain our final expression for $\mathcal{J}_{5}^{\mu}(t,\vx,\vp)$ as
\begin{align}
	\label{eq_G_grad}
&	\mathcal{J}_{5}^{\mu}(t,\vx,\vp)
	=\lim^{\vq\rightarrow0}_{\tilde{\o}\rightarrow0}\frac{i\ve_{u}}{\b}\frac{\partial}{\partial q^{\perp}_\xi} u_\l G_R^{\mu\l\g}(\tilde{\o},q_{\perp},p_{\perp})\pd^{\perp}_{\xi}\b_{\g}\no\\
	&=-\frac{1}{\b\ve_{u}}\frac{\pd n(\ve_{u})}{\pd\ve_{u}} p_\r (\bar{\Delta}^{\xi}_{\sigma}-\frac{p_\perp^{\xi}}{\ve_{u}}u_{\s})\ve^{\r\s\mu(\g}p^{\l)}u_{\l}\partial^{\perp}_{\xi}\b_\g\no\\
	&=n_{0}(1-n_{0})\ve^{\mu\nu\a\b}\Big{\{}p_{\nu}\partial^{\perp}_{\a}\b _{\b}-2  p_\b u_{\nu}\frac{ p^{\g}}{\ve_u}\partial^{\perp}_{(\a}\b_ {\g)}
	\Big{\}}
\end{align}
where $n_0$ is the abbreviation for $n(\ve_u)$. This formula exactly agrees with our findings in Ref.~\cite{Liu:2021uhn}. To make it apparent, we need to split the thermal vorticity effects into spin-Nernst effects and vorticity effects, as discussed in \cite{Liu:2022zxd}. Meanwhile, the differences between our results and the findings in \cite{Becattini:2021iol} are also briefly discussed in Ref.~\cite{Liu:2022zxd}.

To obtain the final polarization of the spin-1/2 particle, we need to normalize the $J^\m_5$ current density with the particle's density, which is essentially $2n_0$. We also need a mass in the denominator to convert the axial current density $J_5^\mu$ into the Pauli-Lubanski spin density. Then, our final expression for the polarization in a covariant form is:
\begin{align}
	\label{eq_pol-half}
	P^\mu(p)=\ddfrac{\int d\Sigma^{\l} p_\l \mathcal{J}^\mu_5(t,\vx,\vp)}{2 m \int d\Sigma^\l p_\l n_0(t,\vx,\vp)}
\end{align}
\subsection{Mode selection: slow and fast modes}
\label{sec_mode}
When we evaluate the integral over $k_0$ and $k'_0$ in Eq.~(\ref{eq_half_longG}) for the full $J^\mu_5$, we will find that it contains four modes, corresponding to the four ``$\pm$" combinations of $(s_{k_0}, s_{k'_0})$ in particle-antiparticle. We will call them particle density mode (pp) (positive, positive), antiparticle density mode for the $-p$ direction (nn) (negative, negative), particle-antiparticle axial vector meson mode (incoming) (pn), and antiparticle-particle anti-axial vector meson (outgoing) mode (np). 

The previous two modes have poles around $\tilde{\o} \sim 0^+$ or $0^-$, so they are slow modes or zero modes. Meanwhile, the axial meson modes have poles around $\tilde{\o} \approx 2\ve_{\vp}$, which are fast modes and can be regarded as virtual axial vector mesons close to the thresholds.

Both slow and fast modes will survive in the slow limits, where we take $\tilde{\o} \rightarrow 0$ before taking $\vq \rightarrow 0$, although there is a factor of $T/\ve_{\vp}$ suppression since it is away from the pole. In the fast limit, $\vq \rightarrow 0$ is taken first before $\tilde{\o} \rightarrow 0$, which makes the slow modes vanish but allows the fast modes to survive as expected. Therefore, if we use the slow limits minus the fast limits, i.e., the ``slow minus fast" scheme as discussed in Ref.~\cite{Liu:2021uhn}, we effectively select the slow modes corresponding to the modes of particle/antiparticle polarization density, which are the modes related to the final observables.

As discussed, the correct modes for the particle's polarization can be extracted if we only integrate the region covering the poles corresponding to $(s_{k_0} = 1, s_{k'_0} = 1)$ (pp) mode. This procedure will select the mode close to $\tilde{\o} \sim 0^+$. To elaborate further, we can check the time structure of $\<J_5^{\mu}\>$ for operator defined in Eq.~(\ref{eq_J5de})
Schematically, for a free field at global equilibrium, its time structure can be represented as
\begin{align}
	\label{eq_J5LSZt}
	&\mathcal{J}_{5}^\mu(t)\propto C_{\rm pp}^\mu\langle a^{\dagger}_{\vk'} a_{\vk}\rangle e^{-i(\ve_{\vk}-\ve_{\vk'})t}+C_{\rm nn}^\mu\langle b^\dagger_{\vk'} b_{\vk}\rangle e^{-i(\ve_{\vk'}-\ve_{\vk})t}\no\\
	&+C_{pn}^\mu\langle a_{\vk} b_{\vk'}\rangle e^{-i(\ve_{\vk}+\ve_{\vk'})t}+C_{np}^\mu\langle a^{\dagger}_{\vk} b^\dagger_{\vk'}\rangle e^{i(\ve_{\vk'}+\ve_{\vk})t}
\end{align}
which is corresponding to the pole structure in Fourier space as
\begin{align}
	\label{eq_J5LSZpole}
	\mathcal{J}_{5}^\mu(\tilde{\o})\propto G^\mu_R(\tilde{\o})\propto &\frac{C_{\rm pp}^\mu\langle a^{\dagger}_{\vk'} a_{\vk}\rangle}{\tilde{\o}+\ve_{\vk'}-\ve_{\vk}}+\frac{C_{\rm nn}^\mu\langle b^{\dagger}_{\vk'} b_{\vk}\rangle}{\tilde{\o}-\ve_{\vk'}+\ve_{\vk}}\no\\
	&+\frac{C_{pn}^\mu\langle a_{\vk} b_{\vk'}\rangle }{\tilde{\o}-\ve_{\vk'}-\ve_{\vk}}+\frac{C_{np}^\mu\langle a^{\dagger}_{\vk} b^\dagger_{\vk'}\rangle}{\tilde{\o}+\ve_{\vk'}+\ve_{\vk}}
\end{align}

In quantum field theory, when we would like extract the modes from a correlation functions, typically we use the LSZ reduction techniques~\cite{Peskin:1995ev}. The most straightforward realization of the LSZ procedure is to multiplying the whole correlation function in the momentum space with the pole, taking it to zero to select the residues out as shown in the below:
\begin{align}
	\label{eq_J5LSZw}
	J^{\mu}_{5,+}(\tilde{\o})=\frac{\lim_{\tilde{\o}\rightarrow\ve_{\vk}-\ve_{\vk'}}(\tilde{\o}+\ve_{\vk'}-\ve_{\vk}) \mathcal{J}_{5}^\mu(\tilde{\o})]}{\tilde{\o}+\ve_{\vk'}-\ve_{\vk}}
\end{align}
Note that the $\tilde{\o}$ structure of $J_5^\mu$ is inherited from the retarded Green function $G_R$. Therefore, the standard LSZ reduction to select out $J^{\mu}_{5,+}$ is equivalent to only selecting the density mode (pp), as done in the previous section. For the off-shell case, it will be more complicated, and the pole selection procedure will be more obscure. However, with the understanding of the physical meanings of the four modes, the selection procedure practiced in this work can still be regarded as a generalized LSZ scheme for off-shell correlation functions.

\section{Vector polarization for spin-1}
\label{sec_spinonevec}
Following the discussion of vector polarization for spin-1/2 particles, we now extend the calculation to vector polarization for spin-1 particles using the same Zubarev response approach. In this section, we will elaborate on the details of the calculation of the vector polarization results that have already briefly reported in Ref.~\cite{Li:2022vmb}. Meanwhile, many aspects of the calculation of vector polarization are intrinsically connected to the tensor polarization/spin alignment calculation. Therefore, obtaining expected results for vector polarization can serve as a benchmark for the correctness of the tensor polarization/spin alignment calculations.

\subsection{Various conserved currents}
\label{sec_current}
The energy-momentum tensor for the Proca field in the literature is sometimes simplified using partial integration methods at the $T^{\mu\nu}$ level~\cite{Serot:1992ti}. However, this is not allowed for our purpose since we will couple $T^{\mu\nu}$ with $\beta^\mu(x)$ rather than some constant inside the integral. Therefore, we will repeat the derivation of $T^{\mu\nu}$ directly from the Proca Lagrangian density of the massive vector boson:
\begin{align}
\label{eq_Ldensity}
\mathcal{L}=-\frac{1}{4}\tensor{F}{^{\mu\nu}}\tensor{F}{_{\mu\nu}}+\frac{1}{2}m^2 V^\mu V_\mu\,,
\end{align}
where field strength is defined as $F^{\mu\nu} = \pd^\mu V^\nu - \pd^\nu V^\mu$. 

The equation of motion (EoM) for the vector fields is $\partial_{\mu} F^{\mu\nu} + m^2 V^{\nu} = 0$. A further derivative with respect to ``$\nu$" generates the constraint equation $\partial_\nu V^{\nu} = 0$, making $V^\mu$ transverse. Using the EoM or Noether's theorem, we can construct a generic conserved current $X^\mu$ as
\begin{align}
	\label{eq_conJ}
	X^{\mu}=\frac{\partial \mathcal{L}}{\partial (\partial_{\mu} V_{\a})}\delta V_{\a}-Y^\mu
\end{align}
$Y^\mu$ is related to the total derivative of $\mathcal{L}$. For the energy-momentum tensor, $\delta V_{\alpha} = \e_{\nu} \pd^\nu V_{\alpha}$, and $Y^{\mu} = \e_{\nu} \eta^{\mu\nu} \mathcal{L}$. Inserting this into Eq.~(\ref{eq_conJ}) and factoring out $\e_{\nu}$, we obtain the canonical energy-momentum tensor as
\begin{align}
	\label{eq_T_can}
	T^{\mu\nu}_C=-F^{\mu\a}\partial^{\nu} V_{\a}-\eta^{\mu\nu}(-\frac{1}{4}F^{\alpha\beta}F_{\alpha\beta}+\frac{1}{2}m^2 V^\a V_\a)
\end{align}


However, the canonical energy-momentum tensor is not symmetric in the ``$\mu\,\nu$" indices. Following our previous work~\cite{Liu:2021uhn}, we will simply choose the symmetric Belinfante form. To proceed, we first obtain the angular momentum tensor, for which we need $\delta V_{\nu} = (x^{\alpha} \pd^{\beta} V_{\nu}-V^{\a}\d^{\b}_{\n}) \o_{\alpha\beta}$ and $Y^\mu = \o_{\alpha\beta} x^{\alpha} \eta^{\beta\mu} \mathcal{L}$, originating from $\delta \mathcal{L} = \o_{\alpha\beta} x^{\alpha} \pd^\beta \mathcal{L} = \o_{\alpha\beta} \pd_{\mu}(x^{\alpha} \eta^{\beta\mu} \mathcal{L})$. Factoring out $\o_{\alpha\beta}$, we obtain the total angular momentum tensor $J^{\mu\alpha\beta}$, the orbital angular momentum $L^{\mu\alpha\beta}$, and the spin tensor $S^{\mu\alpha\beta}$ as
\begin{align}
	\label{eq_J}
	&J^{\mu\alpha\beta}=S^{\mu\a\b}+L^{\mu\a\b},\no\\
	&S^{\mu\a\b}=\tensor{F}{^{\mu\b}}V^{\a}-\tensor{F}{^{\mu\a}}V^{\b},\no\\ &L^{\mu\a\b}=x^\a\tensor{T}{^{\mu \b}}-x^\b\tensor{T}{^{\mu \a}}
\end{align}
With the spin operator, the Belinfante form of the energy-momentum tensor is $T^{\mu\nu} = T^{\mu\nu}_C + (1/2) \pd_{\alpha}[S^{\alpha\mu\nu} - S^{\mu\alpha\nu} - S^{\nu\alpha\mu}]$, which uses the derivative of an anti-symmetric tensor that automatically satisfies the conservation law to cancel the anti-symmetric part of $T_C^{\mu\nu}$ (note $\pd_\alpha S^{\alpha\mu\nu} = -2T_C^{[\mu\nu]}$ since $\pd_{\alpha} J^{\alpha\mu\nu} = 0$). We will suppress the sub-index and denote the Belinfante form of the energy-momentum tensor as $T^{\mu\nu}$
\begin{align}
	\label{eq_Tmn_vec}
	T^{\mu\nu}=&-\tensor{F}{^{\mu}^{\alpha}}\tensor{F}{^{\nu}_{\alpha}}+m^2V^{\mu}V^{\nu}\no\\
	&-(-\frac{1}{4}F^{\alpha\beta}F_{\alpha\beta}+\frac{1}{2}m^2V^\a V_\a)\eta^{\mu\nu}\;,
\end{align}
which is the energy momentum tensor used in following sections.

\subsection{Main formalism}
Similar to the procedures done in the literature~\cite{Liu:2021uhn}, we first define the axial vector current. As $\mathcal{J}^\mu_5\propto\ve^{\mu \nu\a\b}S_{\nu\a\b} $ in the spin-1/2 case, we construct the axial vector current in a similar way as $\mathcal{J}^{\mu}_{5}$ as
\begin{align}
	\label{eq_J^5_nohermi}
	&\hat{\mathcal{J}}^{\mu}_{5}(x)=-\frac{1}{2}\ve^{\mu \nu\a\b}S_{\nu\a\b}=2\ve^{\mu \nu\a\b}(\pd_\nu V_\a(x)) V_{\b}(x)
\end{align}
To make the above $\hat{\mathcal{J}}^{\mu}_5$ Hermitian, we use the above equation plus its conjugate as
\begin{align}
	\label{eq_J^5}
	&\hat{\mathcal{J}}^{\mu}_{5}(x)=\ve^{\nu \mu\a\b}V_\a(x) \overset{\leftrightarrow}{\pd_{\nu}}  V_{\b}(x)
\end{align}
The overall factor can be fixed when evaluating the spin at particle rest frame. Meanwhile, this definition will be consistent with the $m\Tr[S^{\mu}\rho]$ as discussed in Sec.~\ref{sec_density}. 

To proceed, we further define Wigner function as
\begin{align}
	\label{eq_Wdefine}
W^{\mu\nu}(x,\vp)&\equiv2\varepsilon_{\vp} \int {d^3 \vy}e^{-i \vp\cdot \vy}\langle V^{\mu}(t, \vx_-)V^{\nu}(t,\vx_+)\rangle
\end{align}
where $\vx_{\pm}=\vx\pm\vy/2$ , and the \textbf{extra factor ``2"} mentioned in Ref.~\cite{Li:2022vmb} is absorbed into this definition. With the Wigner function, Eq.~(\ref{eq_J^5}) can be represented as
\begin{align}
	\label{eq_J5_pspace}
	&\mathcal{J}^{\mu}_{5}(t,\vx,\vp)\no\\
	&=\ve^{\nu \mu\a\b}\ve_{\vp}  \int d^3 \vy e^{-i\vp\cdot \vy}\langle V_{\a}(t,\vx_{-})\overset{\leftrightarrow}{\pd_{\nu}^{x}} V_{\b}(t,\vx_{+})\rangle
	\no\\
	&=\ve^{\nu \mu\a\b} (-ip_\nu)W_{\a\b}(x,\vp)+o(\pd)\,.
\end{align}
As shown, accurate to leading order in gradients (or $\vq$), the derivatives can be factored out as the momentum $p$. The relation in the last line will manifest in the concrete calculation later. Therefore, we can calculate the Wigner function $W_{\a\b}(x,\vp)$ separately. 

Additionally, according to Eq.~(\ref{eq_nd_final}) and Eq.~(\ref{eq_od_time}), $W_{\alpha\beta}(t, \vx, \vp)$ will be related to the retarded Green function $G^{\mu\nu\l\g}_R(t-t',\vx,\vx',\vy)$ as defined in the following:
\begin{align}
	\label{eq_def_Gr}
	&G^{\mu\nu\l\g}_R(t-t',\vx,\vx',\vy)\equiv\int \frac{d\omega}{2\pi} \frac{d^3\vq}{(2\pi)^3} \frac{d^3\vp}{(2\pi)^3}e^{-i\omega\cdot(t-t')}
	\no\\
	&\hspace{3cm}\times e^{i\vq\cdot(\vx-\vx')}e^{i\vp\cdot\vy}G^{\mu\nu \l\g}_{R}(\omega,\vq,\vp)
	\no\\
	&=-i\theta(t-t')\langle[ V^\mu(t,\vx^{-}) V^{\nu}(t,\vx^{+}),T^{\l\g}(t',\vx')]\rangle\,.
\end{align}
Before we discuss how to exactly relate $G^{\mu\nu\lambda\gamma}_R$ to $W_{\alpha\beta}(t, \vx, \vp)$ as expressed in Eq.~(\ref{eq_wum_anti}) or discussed in Sec.~\ref{sec_spinoneten}, we first explain how to calculate it using the imaginary time formalism, as done in the previous section. 

As explained in Sec.~\ref{sec_conti}, the $G^{\mu\nu\lambda\gamma}_R$  will be directly related to the imaginary/Euclidean time Green function:
\begin{align}
	\label{eq_def_Gr_E}
&G^{\mu\nu\l\g}(\tau-\tau',\vx,\vx',\vy)=\no\\
&-\langle T_\tau [ V^\mu(\tau,\vx_{-}) V^\nu(\tau,\vx_{+})T^{\l\g}(\tau',\vx')]\rangle
\end{align}
In evaluating Eq.~(\ref{eq_def_Gr_E}) with Eq.~(\ref{eq_Tmn_vec}), the expression $\langle T_{\t}[ V^\m V^\n F^{\l\a}F^{\g\b}]\rangle$ will be useful, which is
\begin{align}
	\label{eq_contract_FF}
&\langle T_\t [  V^\mu(\tau,\vx_{-}) V^\nu(\tau,\vx_{+})F^{\l\a}(\tau',\vx')F^{\g\b}(\tau',\vx')]\rangle\no\\
&=[\tilde{\pd}^{\prime\l}D^{\n\a}(r_{\rm o})-\tilde{\pd}^{\prime\a}D^{\n\l}(r_{\rm o})][\tilde{\pd}^{\prime\g}D^{\b\m}(r_{\rm i})-\tilde{\pd}^{\prime\b}D^{\g\m}(r_{\rm i})]\no\\
&+[\g\leftrightarrow \l,\a\leftrightarrow \b]
\end{align}
where $r_{\rm o}=(\tau-\tau',\vx_{+}-\vx')$,$r_{\rm i}=(\tau'-\tau,\vx'-\vx_{-})$ and $[\g\leftrightarrow \l,\a\leftrightarrow \b]$ means the expression with indices exchanged as indicated. The $D^{\mu\nu}$ is the full/dressed propagator for a vector particle. As explained in the text around Eq.~(\ref{eq_j5_SS}) to Eq.~(\ref{eq_S_deriv}), this $\tilde{\pd}$ excludes the $\d(\t)$ generated from the derivative of the $\theta$ function in time-ordered products. 

With Eq.~(\ref{eq_contract_FF}), the correlation function at the leading order of the skeleton expansion (see Sec.~\ref{sec_skeleton}) can be expressed as
\begin{align}
	\label{eq_Gim_t0i2}
	&G^{\mu\nu \l \g}(\tau-\tau',\vx,\vx', \vy)=\Big\{[\tilde{\pd}^{\prime\l}D^{\n\a}(r_{\rm o})-\tilde{\pd}^{\prime\a}D^{\n\l}(r_{\rm o})]\no\\
	&\times[\tilde{\pd}^{\prime\g}D^{\m}_{\a}(r_{\rm i})-\tilde{\pd}'_{\a}D^{\g\m}(r_{\rm i})]-m^2 D^{\nu \l}(r_{\rm o}) D^{\mu \g}(r_i)\no\\
	&-\big{\{}\frac{1}{4}[\tilde{\pd}^{\prime\a}D^{\n\b}(r_{\rm o})-\tilde{\pd}^{\prime\b}D^{\n\a}(r_{\rm o})][\tilde{\pd}^{\prime}_{\a}D^{\m}_{\b}(r_{\rm i})-\tilde{\pd}'_{\b}D^{\m}_{\a}(r_{\rm i})]\no\\
	&
	-\frac{1}{2}m^2 D^{\nu \a}(r) D^{\mu}_{\a}(r')\big{\}}\eta^{\l\g}\Big\}+\Big\{\g\leftrightarrow \l\Big\}
\end{align}
where $\Big\{\g\leftrightarrow \l\Big\}$ means the expression with $\g$ and $\l$ exchanged.

With the spectral function of the vector meson defined as
\begin{align}
	\label{eq_spec}
&A^{\mu\nu}=\sum_{a=L,T}\Delta^{\mu\nu}_{a}A_{a}, \,A_{a}=\frac{1}{\pi}\text{Im }\frac{-1}{p^2-m^2-\Pi_a}.
\end{align}
The full $\Delta^{\mu\nu}$, longitudinal $ \Delta_{L}$ and transverse projector $ \Delta_{T} $ are $\Delta^{\mu\nu}=-\eta^{\mu\nu}+ p^{\mu}p^{\nu}/p^2$, $ \Delta^{\mu\nu}_{L}=v^\mu v^\nu/(-v^2)$ and
$ \Delta^{\mu\nu}_{T}=\Delta^{\mu\nu}-\Delta^{\mu\nu}_{L}$ respectively, where
$v^\mu=\Delta^{\mu\nu}u_\nu$ is the projected flow velocity with respect to $p$.

With the spectral function defined above, the propagator can be obtained by analytical continuation to imaginary time as
\begin{align}
	\label{eq_Dim}
	D^{\mu\nu}(\tau,\vx)&=
	\frac{1}{\beta}\sum_{\o_n}\int \frac{d^3\vk}{(2\pi)^3}\int dk_0 \frac{A^{\mu\nu}(k_0,\vk)}{i\o_n-\o}e^{-i\omega_n\tau +i\vk\cdot\vx}
	\no\\
	&=\frac{1}{\beta}\sum_{\o_n}\int \frac{d^4k}{(2\pi)^3} \frac{A^{\mu\nu}(k)}{i\o_n-\o}e^{-i\omega_n\tau +i\vk\cdot\vx}\,.
\end{align}
Note that due to the convention used in the spectral function, it is $(2\pi)^3$ rather than $(2\pi)^4$ in the denominator. 
Similarly to the spin-1/2 case, the $\tilde{\pd}$ derivatives of it can be expressed as
\begin{align}
\tilde{\pd}^\l D^{\mu\nu}=\frac{1}{\beta}\sum_{\o_n}\int \frac{d^4k}{(2\pi)^3} \frac{-i k^{\l} A^{\mu\nu}(k)}{i\o_n-\o}e^{-i\omega_n\tau +i\vk\cdot\vx}
\end{align}
Note that the derivative $\tilde{\pd}$ excludes the $\delta(\tau)$, as discussed before. Therefore, its overall effect is to generate an internal $-ik$, similar to it in real space.

With everything prepared, we can express the retarded Green function $G_R^{\m\n\l\g}(\tau-\tau^\prime,\vx,\vy,\vx')$ in terms of the spectral functions as
\begin{widetext}
	\begin{align}
		\label{eq_one_longG}
		&G_R^{\mu\nu\l\gamma}(\tau-\tau^\prime,\vx,\vy,\vx')=
		\frac{1}{\beta}\sum_{\omega_n}\frac{1}{\b}\sum_{\nu_n}\int\frac{d^4 k}{(2\pi)^3}\int\frac{d^4 k^\prime}{(2\pi)^3} e^{-i(\o_n+\nu_n)(\tau-\tau^\prime)+i\vk\cdot(\vx+\frac{\vy}{2}-\vx')} e^{-i\nu_n(\tau^\prime-\tau)+i\vk^\prime\cdot(-\vx+\frac{\vy}{2}+\vx')}\no\\
		&\times\frac{1}{i\o_n+i\nu_n-k_0}\frac{1}{i\nu_n-k_0'}\Big\{[k^{\l}A^{\n\a}(k)-k^{\a}A^{\n\l}(k)][k^{\prime\g}A^{\m}_{\a}(k')-k'_{\a}A^{\g\m}(k')]-m^2 A^{\nu \l}(k) A^{\mu \g}(k')\no\\
		&-\big{\{}\frac{1}{4}[k^{\a}A^{\n\b}(k)-k^{\b}A^{\n\a}(k)][k^{\prime}_{\a}A^{\m}_{\b}(k')-k'_{\b}A^{\m}_{\a}(k')]-\frac{1}{2}m^2 A^{\nu \a}(k) A^{\mu}_{\a}(k')\big{\}}\eta^{\l\g}\Big\}+\Big\{\g\leftrightarrow \l\Big\}
	\end{align}
\end{widetext}
where the $k'=(k'_0,\vp-\vq/2)$ and $k=(k_0,\vp+\vq/2)$.
We can use Eq.~(\ref{eq_contour_int}) to carry out the frequency sum. Then, we analytically continue it to $i\o_n \rightarrow \o + i\e$. The retarded Green function defined in momentum space, $G_R^{\m\n\l\g}(\o, \vp, \vq)$, as explained in Eq.~(\ref{eq_def_Gr}), can be organized as follows:
\begin{align}
	\label{eq_one_Gp}
	&G_R^{\mu\nu\l\gamma}(\o,\vp,\vq)=-\int dk_0 dk_0' \frac{n(k_0')-n(k_0)}{\o+k_0'-k_0+i\e}I^{\mu\nu\l\g}
\end{align}
with the tensor part $I^{\mu\nu\l\g}$ as
\begin{align}
	\label{eq_Imnlg}
	&I^{\mu\nu\l\g}=(k^{\l}k'^{\g}+k^{\g}k'^{\l})A^{\n\a}(k)A^{\m}_{\a}(k')\no\\
	&-[k^{\prime\g}A^{\n\l}(k)+k^{\prime\l}A^{\n\g}(k)]k^{\a}A^{\m}_{\a}(k')\no\\
	&-[k^{\l}A^{\g\m}(k')+k^{\g}A^{\l\m}(k')]k'_{\a}A^{\n\a}(k)\no\\
	&+(k\cdot k'-m^2) [A^{\nu \l}(k) A^{\mu \g}(k')+A^{\nu \g}(k) A^{\mu \l}(k')]\no\\
	&-\eta^{\l\g}[(k\cdot k'-m^2 )\eta_{\a\b}-k_{\b}k'_{\a}]
	A^{\nu \a}(k) A^{\mu\b}(k')
\end{align}
Eq.~(\ref{eq_one_Gp}) and Eq.~(\ref{eq_Imnlg}) can be used in the evaluation for both vector polarization and tensor polarization/spin alignment~\cite{Li:2022vmb}. In calculating the vector polarization, we follow what is done in the spin-1/2 case by using the spectral functions for free particles to illustrate the leading effects, where this free spectral function is
\begin{align}
\label{eq_one_freespec}
A^{\mu\nu}(p)&=\Delta^{\mu\nu}(p)s_{p_0}\delta (p_0^2-\varepsilon^2_{\vp})
\end{align}

Taking the density (pp) modes (see Sec.~\ref{sec_mode}), Eq.~(\ref{eq_one_Gp}) can be reduced to
\begin{align}
	\label{eq_one_PGp}
	G_R^{\mu\nu\l\gamma}(\o,\vq,\vp)=-\frac{1}{4\ve_{\vk}\ve_{\vk'}} \frac{n(\ve_{\vk'})-n(\ve_{\vk})}{\ve_{\vk'}-\ve_{\vk}}
	I^{\mu\nu\l\g}_{A\rightarrow\Delta}
\end{align}
$I^{\m\n\l\g}_{A\rightarrow\Delta}$ is just a short notation for replacing $A^{\mu\nu}$ in $I^{\mu\nu\l\g}$ in Eq.~(\ref{eq_Imnlg}) with the projectors $\Delta^{\mu\nu}$ defined after Eq.~(\ref{eq_spec}).

Expanding $G_R^{\mu\nu\l\gamma}(\o,\vq,\vp)$ to leading order in $\vq$ and taking the anti-symmetric part of the ``$\mu, \nu$" indices, we have
\begin{align}
	\label{eq_one_PGpexpand}
	&G_R^{[\mu\nu]\l\gamma}(0,\vq,\vp)\no\\
    &=-\frac{1}{4\ve_{u}^2}\frac{\pd n(\ve_u)}{\pd\ve_u}
	[2p^{\l}p^{\g}\frac{q^{\a}p^{[\nu}\Delta^{\mu]}_{\a}(p)}{m^2}-4p^{(\g}\Delta^{\l)[\n}\Delta^{\mu]}_\a q^\a]\no\\
	&=\frac{4}{4\ve_{u}^2}\frac{\pd n(\ve_u)}{\pd\ve_u}
	[ p^{\l}p^{\g}\frac{p^{[\n}q^{\m]}}{2m^2}+p^{(\g}\eta^{\l)[\n}q^{\mu]}-p^{(\g}p^{\l)}\frac{p^{[\n}q^{\mu]}}{m^2}]\no\\
	&=-\frac{1}{\ve_{u}^2}\beta n_0(1+n_0)[q^{[\m}\eta^{\n](\l}p^{\g)}-\frac{1}{2} p^{(\g}p^{\l)}\frac{p^{[\n}q^{\mu]}}{m^2}]\,.
\end{align}
The first term in Eq.~(\ref{eq_one_PGpexpand}) comes from the first term in Eq.~(\ref{eq_Imnlg}), while the second term comes from the second and third terms in Eq.~(\ref{eq_Imnlg}). For the on-shell case and leading order in $\vq$, $q = (\vq \cdot \vp / \ve_{\vp}, \vq)$, so that $q \cdot p = 0$, and this identity is useful in the derivation of Eq.~(\ref{eq_one_PGpexpand}).

With Eq.~(\ref{eq_def_Gr}) and Eq.~(\ref{eq_nd_final}), we can express the anti-symmetric part of the Wigner function as
\begin{align}
	\label{eq_wum_anti}
	\delta W_{[\a\b]}=\lim^{\vq\rightarrow0}_{\o\rightarrow0}\frac{2i\ve_{u}}{\b}\frac{\partial}{\partial q^{\perp}_\xi} [u^{\l}G^{R}_{[\a\b]\l\g}(\o,\vq,\vp)]\pd^{\perp}_{\xi}\b^{\g}\,.
\end{align}
With Eq.~(\ref{eq_J5_pspace}), Eq.~(\ref{eq_wum_anti}), and the reiterated Eq.~(\ref{eq_dperp}) as below:
\begin{align}
	\label{eq_dperp2}
	\frac{\pd}{\pd q^\perp_{\xi}}(q^{\perp}_{\a}+(q\cdot u)u_{\a})=\bar{\Delta}^{\xi}_{\a}-\frac{p_\perp^{\xi}}{\ve_{u}}u_{\a}\;,
\end{align}
the axial current phase space density 	$\mathcal{J}_{5}^{\mu}(t,\vx,\vp)$ can be expressed as
\begin{align}
	\label{eq_J5-one}
	&	\mathcal{J}_{5}^{\mu}(t,\vx,\vp)\no\\
	&=\ve^{\nu \mu\a\b} (-ip_\nu)\lim^{\vq\rightarrow0}_{\o\rightarrow0}\frac{2i\ve_{u}}{\b}\frac{\partial}{\partial q^{\perp}_\xi} u^{\l}G^{R}_{\a\b\l\g}(\o,\vq,\vp)\pd^{\perp}_{\xi}\b^{\g}\no\\
	&=n_{0}(1+n_{0}) \frac{-1}{\ve_{u}}p_\n (\bar{\Delta}^{\xi}_{\a}-\frac{p_\perp^{\xi}}{\ve_u}u_{\a})\ve^{\n\m\a(\l}p^{\g)}u_{\l}\partial^{\perp}_{\xi}\b_\g\no\\
	&=n_{0}(1+n_{0})\ve^{\mu\nu\a\b}\Big{\{}p_{\nu}\partial^{\perp}_{\a}\b _{\b}-2  p_\b u_{\nu}\frac{ p^{\g}}{\ve_u}\partial^{\perp}_{(\a}\b_ {\g)}
	\Big{\}}.
\end{align}
 The second term  in the last line of Eq.~(\ref{eq_one_PGpexpand})  does not contribute to the result. Note that $\mathcal{J}^\mu_5(t, \vx, \vp)$ here does not depend on the mass. Thus, it can be smoothly connected to the massless case, for example, photons. In this sense, photons should have shear-induced polarization~\cite{Liu:2021uhn,Fu:2021pok} in a thermalized medium.

For massive vector bosons, the polarization is schematically $\mathcal{J}^\mu_5(t, \vx, \vp) / (m \times 3 n_0)$ in the Cooper-Frye-like formula. More precisely, the polarization can be expressed as
\begin{align}
	P^\mu(p)=\ddfrac{\int d\Sigma^{\l} p_\l \mathcal{J}^\mu_5(t,\vx,\vp)}{3 m \int d\Sigma^\l p_\l n_0(t,\vx,\vp)}
\end{align}
This is $4/3$ times the polarization form of a spin-1/2 particle, as spin-1 particles have 2 times more angular momentum but $3/2$ more states. For photons, we should replace $m$ by $\ve_{u}$ and $3$ by $2$ for their polarization.

\subsection{For the covariance of density matrices}
\label{sec_density}
There is a discussion in the literature regarding different definitions of the spin density matrices~\cite{Gao:2023wwo}, which differ from those used in Ref.~\cite{Becattini:2020sww}. Meanwhile, the different definitions will lead to different results. We realized this subtlety when calculating the vector polarization in our work~\cite{Li:2022vmb}. In this section,  we will try to clarify it use the vector boson as an example.

The first definition is the one used in Refs.~\cite{Hattori:2020gqh,Becattini:2020sww,Li:2022vmb}, in which the (density of) density matrix is
\begin{align}
	\label{eq_rho_def1}
	\varrho^{(w)}_{ s's}(t,\vx,\vp)= \e_{\m,s'}(\vp)W^{\m\n}(t,\vx,\vp)\e^{*}_{\nu, s}(\vp)
\end{align}
On the other hand, the second one is explained in Ref.~\cite{Gao:2023wwo}, which is
\begin{align}
	\label{eq_rho_def2}
	&\varrho^{(a)}_{ s's}(t,\vx,\vp)=\int \frac{d^3\vq}{(2\pi)^3} e^{i\vq\cdot \vx}\no\\
	&\times\e_{\m,s'}(\vp-\vq/2)W^{\m\n}(t,\vq,\vp)\e^{*}_{\nu, s}(\vp+\vq/2)\no\\
	&=\int \frac{d^3\vq}{(2\pi)^3} e^{i\vq\cdot \vx}\langle a^{\dagger}_{\vp-\vq/2,s'}a_{\vp+\vq/2,s}\rangle+o(\pd)
\end{align}
To obtain third line from the second line, we need $\ve_{\vp} / \sqrt{\ve_{\vk} \ve_{\vk'}} \sim 1$ in the small $\vq$ limit. 

Superficially, both the Wigner-based definition $\varrho^{w}$ in Eq.~(\ref{eq_rho_def1}) and the ``$a^\dagger a$"-based definition $\varrho^{(a)}$ in Eq.~(\ref{eq_rho_def2}) are justified. In the following, we will discuss their transformation properties under Lorentz boosts, where the differences and subtleties appear.

First, we use the notation $\Lambda$ to define a boost that takes $p = (\ve_{\vp}, \vp)$ to $p' = (\ve_{\vp'}, \vp')$, and $x = (t, \vx)$ to $x'$. On the other hand, $L(p)$ is the standard boost, taking $  (m, \bm{0})$ to $p$ (see Ref.~\cite{Weinberg:1995mt}). Then, the little group $W$ denotes the ``rotations" contained in $\Lambda$ and is represented by $W(\La, p) = L^{-1}(p') \La L(p)$, which is an element of the SO(3) group. $D^{(1)}_{s s'}(W(\Lambda,p))\equiv D^{(1)}_{s s'}(\Lambda,\vp)$  is the representation that transforms the spin indices of a spin-1 particle at momentum $p$ under the boost $\Lambda$. The transformation of these indices will have the effect that $D^{(1)}_{\s' s' }(\La,\vp)\e_{\m',\s'}(\vp')=\La_{\m'}^{\;\;\g}\e_{\g,s'}(\vp)$ (repeated indices imply summation).

With all the preparation above, $\varrho^{(w)}_{s's}$ defined in Eq.~(\ref{eq_rho_def1}) will transform as
\begin{align}
	\label{eq_rho_def1trans}
	&\varrho^{(w)}_{s's}(x,\vp)\no \\
    &=\Lambda_{\m'}^{\;\;\g}\e_{\g,\s'}(\vp)\Lambda_{\n'}^{\;\;\l}\e^{*}_{\l,\s}(\vp)\Lambda^{\mu'}_{\;\;\mu}\Lambda^{\nu'}_{\;\;\nu}W^{\mu\nu}(x,\vp)
	\no\\
	&=D^{(1)}_{\s' s' }(\Lambda,\vp)\e_{\m',\s'}(\vp')W^{\mu'\nu'}(\Lambda ^{-1}x',\vp',u'...l')\e^{*}_{\mu',\s}(\vp')\no\\
	&\quad\times D^{(1)*}_{\s s }(\Lambda,\vp)\no\\
	&=D^{(1)}_{\s' s' }(\Lambda,\vp)\varrho^{(w)}_{\s'\s}(\Lambda^{-1}x',\vp',u'...l')D^{(1)*}_{\s s }(\Lambda,\vp)
\end{align}
Here, $u' = \La u$ and $l' = \La l$ represent the boosted generic vectors inside the expressions of the Wigner function or density matrices, in addition to $x$ and $p$. This is what we expect for a tensor with spin indices $s, s'$ transformed under the little group of a Lorentz transformation.

For the second definition, i.e., $\varrho^{(a)}_{s's}$ in Eq.~(\ref{eq_rho_def2}), the transformation is approximately shown as below:
\begin{align}
	\label{eq_rho_def2trans}
	\varrho^{(a)}_{s's}(x,\vp)&\approx\int \frac{d^3\vq'}{(2\pi)^3} e^{i\vq'\cdot \vx'} D^{(1)}_{\s' s' }(\Lambda,\vp-\frac{\vq}{2})\no\\
	&\times\langle a^{\dagger}_{\vp'-\vq'/2,\s'}a_{\vp'+\vq'/2,\s}\rangle D^{(1)*}_{\s s }(\Lambda,\vp+\frac{\vq}{2})
\end{align}
The approximation also simply means neglecting some higher-order contributions in $\vq$. As shown, for scalar particles, there is no problem with this definition. However, for particles with nonzero spins, the two definitions will differ. The transformation in Eq.~(\ref{eq_rho_def2trans}) does not seem to have a simple factorization, as shown in Eq.~(\ref{eq_rho_def1trans}). Therefore, to maintain the covariance of the final operator average, some effort is required, which will be discussed in the following.

We will first discuss the commonly used spin operator $S^\mu_{ss'}(\vp)$ with spin indices, which is expressed as
\begin{align}
	S_{ss'}^\mu(\vp)=\frac{1}{m}\varepsilon^{\n\m\a\b}(-ip_\nu)\e_{\b,s}(\vp)\e^{*}_{\a,s'}(\vp)
\end{align}
In the particle rest frame, $\e^{\a}_{\pm}=\mp(0,1,\pm i,0)/\sqrt{2}$, $\e^{\a}_{0}=(0,0,0,1)$, and $\e^{\a}_{s}(\vp)$  and $\e^{\a}_{s}(\bm{0})$ are connected by $L(p)$. For $S_{ss'}^\mu(\vp)$ in the particle rest frame, $S^0$ is zero, and $S^{i}$ are the familiar representations of the rotational generator for a spin-one particle.
\begin{align}
	\bm{S}=
	\begin{pmatrix}
		\frac{1}{\sqrt{2}}{	\setlength{\tabcolsep}{25pt}
			\begin{pmatrix}
				0 & \;1 & \;0\\
				1 & \;0 & \;1\\
				0 & \;1 & \;0
		\end{pmatrix}},
		\frac{1}{\sqrt{2}}\begin{pmatrix}
			0 & -i & \;0\\
			i & \;0 & -i\\
			0 & \;i & \;0
		\end{pmatrix}
		,
		\begin{pmatrix}
			1 & \;0 & \;0\\
			0 & \;0 & \;0\\
			0 & \;0 & -1
		\end{pmatrix}
	\end{pmatrix}
\end{align}
The $S^\mu(\vp)$ and $S^\mu(0)$ are also connected by $L(p)$. However, if we further transform $S^\mu_{ss'}(\vp)$ by the boost $\La$, the transformation law is
\begin{align}
	\label{eq_Srhotrans}
	&\Lambda^{\mu}_{\;\;\nu} S^\n_{ss'}(\vp)=D^{(1)}_{\s s }(\Lambda,\vp)S^\mu_{\s\s'}(\vp')D^{(1)*}_{\s' s' }(\Lambda,\vp)
\end{align}
As shown, the transformation for a general $S^\mu$ also includes a little group rotation on the spin indices. However, if we trace this $S^\mu$ with the density matrices $\varrho^{w}$, the rotation will cancel each other, as shown below:
\begin{align}
\text{Boost }\text{Tr}[S^\mu\varrho^{(w)}]&=\Lambda^{\mu}_{\;\;\nu} \text{Tr}[D^{*-1} S^\nu D^{*}D^{*-1}\varrho^{(w)} D^{*}]\no\\
&=\Lambda^{\mu}_{\;\;\nu} \text{Tr}[S^\nu\varrho^{(w)}]\;,
\end{align}
which finally becomes the expected transformation for a boost of a general vector.

For the second definition of $\varrho^{(a)}$ in Eq.~(\ref{eq_rho_def2}), if we want to achieve covariance for the operator average, we need to define a different $S^\mu$ operator $S^\mu(\vp, \vq)$ at $\vp, \vq$ space. Schematically, we should have
\begin{align}
	&S_{ss'}^{{(a)\mu}}(\vp,\vq)=\frac{1}{m}\varepsilon^{\n\m\a\b}(-ip_\nu)\e_{\b,s}(\vp+\vq/2)\e^{*}_{\a,s'}(\vp-\vq/2)\no\\
&S_{ss'}^{(a)\mu}(\vx-\vx',\vp)=\int \frac{d^3\vq }{(2\pi)^3}e^{i\vq\cdot( \vx-\vx')}S^{(a)\mu}_{ss'}(\vp,\vq)
\no\\
&S^{(a)\mu}(\vx,\vp)=\int d^{3}\vx'\text{Tr}\{S^{(a)\mu}(\vx-\vx',\vp)\varrho^{(a)}(t,\vx',\vp)\}
\end{align}
Basically, we need a different and non-local definition $S^{(a)\mu}_{ss'}(\vx - \vx', \vp)$ for the spin matrix in the spinor space. Note that, in this case, for small $\vq$ limits, the definition of $S^{\mu}$ includes a Berry connection contribution as shown below
\begin{align}
&S_{ss'}^{(a)\mu}(\vp,\vq)\no\\
&=S_{ss'}^\mu(p)+\frac{1}{2m}\varepsilon^{\n\m\a\b}(-ip_\nu)\e^{*}_{\a,s'}(\vp)\vq^{i}\overset{\leftrightarrow}{\frac{\pd}{\pd{\vp^{i}}}}\e_{\b,s}(\vp)\,.
\end{align}
Therefore, if we use different definitions of density matrices, we are also required to use different definitions of the operator to match the density matrix.

\section{Tensor polarization and alignment for vector boson}
\label{sec_spinoneten}
For the main discussions on spin alignment, we refer to our previous work~\cite{Li:2022vmb}. Since the paper~\cite{Li:2022vmb} is still under review and awaiting publication, it is better to avoid excessive overlap that could affect its publication. In this section, we will mainly focus on aspects that were not thoroughly discussed there.

We will begin with the definition of the tensor polarization and spin alignment based on the discussions in the previous section. Unlike the vector polarization, which is related to the anti-symmetric part of the Wigner function, the tensor polarization will be related to the symmetric and traceless part of the Wigner function with proper projection. This projection can be implemented by the shell projector defined as
\begin{align}
	\tilde{\D}_{\m\n}=-\eta^{\m\n}+\frac{\tilde{p}^{\m}\tilde{p}^{\n}}{m^2}\;.
\end{align}
With this projector, the scalar part of the Wigner function is defined as
\begin{align}
	\sS\equiv W^{\mu\nu}\tilde{\Delta}_{\mu\nu} \approx 3n(\ve_u), \delta S=\delta W^{\mu\nu}\tilde{\Delta}_{\mu\nu} 
\end{align}
where $\delta W^{\mu\nu}$ is the Wigner function deviating from global equilibrium. With all the definitions above, the tensor polarization can be expressed as
\begin{align}
	\label{eq_Tdefine}
	&\mathcal{T}^{\mu\nu}\equiv \delta W^{\langle\mu\nu\rangle}\equiv\tilde\Delta^{\langle\mu}_{\;\l}\tilde\Delta^{\nu\rangle}_{\;\g}\delta W^{(\l\g)}\no\\
	&\tilde{\D}^{\langle\m}_{\;\l}\tilde{\D}^{\nu\rangle}_{\;\g}= \tilde{\D}^{(\m}_{\;\l}\tilde{\D}^{\nu)}_{\;\g}-\frac{1}{3}\tilde{\D}^{\m\n}\tilde{\D}_{\l\g}
\end{align}
With the tensor polarization defined above, the spin alignment $\rho_{00}$ can be expressed as
\begin{align}
	\label{eq_rho-nT}
	\delta \rho_{00}(\hat{n}_{\text{pr}},\vp)=\frac{\int d\Sigma^{\l}p_{\l}\, \sT^{\mu\nu}(x,\vp)\hn_{\mu}(\vp)\hn_{\nu}(\vp)}{d\Sigma^{\l}p_{\l} 3 n(x,\vp) }\,.
\end{align}
$\hat n^\mu$ is a shorthand of $\epsilon^\mu_{s=0}$,  where $\hat{n}^{\mu}=[\Lambda(\mathbf p).\hat{n}_{\text{pr}}]^{\mu}$. The $\hat{n}_{\text{pr}}$ is the polarization vector $\epsilon^\mu_{s=0,\text{pr}}$ in the particle rest frame.
\subsection{Non-dissipative contribution at $O(\pd)$: vanishing}
With the notations discussed above, we can express the non-dissipative part of the tensor polarization following similar procedures in Eq.~(\ref{eq_wum_anti}), which is:
\begin{align}
	\label{eq_tensor_nd}
	\sT_{\rm nd}^{\m\n}(x,\vp)=\lim_{\o\rightarrow0}^{\vq\rightarrow 0}\frac{2i\ve_{u}}{\b}\frac{\partial}{\partial q^{\perp}_\xi} [u_{\l}G^{\langle\m\n\rangle\l\g}_R(\o,\vq,\vp)]\pd^{\perp}_{\xi}\b_{\g}
\end{align}

As implicitly indicated in Ref.~\cite{Li:2022vmb}, the contribution from the non-dissipative part is zero at leading order in gradients. In this section, we provide an explicit calculation of it.


With the energy-momentum tensor given in this work, we will explicitly prove that at leading skeleton order (see Sec.~\ref{sec_skeleton}) and leading hydrodynamic gradient, it is zero, which means the conclusion holds even for off-shell spectral functions. The key to proving this is to check the symmetric part of the tensor $I^{\mu\nu\l\g}(k,k')$ in Eq.~(\ref{eq_Imnlg}), which is
\begin{align}
	\label{eq_Imnlg2}
	&I^{(\mu\nu)\l\g}(k,k')=(2k^{(\l}k'^{\g)})A^{\a(\n}(k)A^{\m)\b}(k')\eta_{\a\b}\no\\
	&-[2k^{\prime(\g}A^{\l)(\n}(k)]A^{\m)}_{\a}(k')k^{\a}-[2k^{(\l}A^{\g)(\m}(k')]A^{\n)}_{\a}(k)k'^{\a}\no\\
	&+(k\cdot k'-m^2) [2A^{(\nu\l }(k) A^{\mu \g)}(k')\no\\
	&-\eta^{\l\g}[(k\cdot k'-m^2 )\eta_{\a\b}-k_{\b}k'_{\a}]
	A^{ \a(\nu}(k) A^{\mu)\b}(k')\no\\
	&\equiv I^{(\mu\nu)\l\g}(k',k)
\end{align}
We find that  $I^{(\mu\nu)\l\g}(k,k')$  has symmetry under the exchange of $k$ and $k'$, which simply means changing ``$\vq$" to ``$-\vq$". Therefore, we have
\begin{align}
	G_R^{\mu\nu\l\gamma}(0,-\vq,\vp)&=-\int dk_0 dk_0' \frac{n(k_0)-n(k_0')}{k_0-k_0'}I^{\mu\nu\l\g}(k',k)\no\\
	&\equiv G_R^{\mu\nu\l\gamma}(0,\vq,\vp)
\end{align}
The above relation suggests that for this energy-momentum tensor at leading skeleton loop, since $G_R$ is even in $\vq$, we have $(\partial/\partial q^{\perp}_\xi) [u^{\lambda} G_{(\alpha\beta)\lambda\gamma}(0, \vq, \vp)] \equiv 0$, so that
\begin{align}
	\delta W_{\rm nd}^{(\a\b)}\equiv 0, \quad
	\sT_{\rm nd}^{\a\b}\equiv \delta W_{\rm nd}^{\langle\a\b\rangle}=0
\end{align}



As discussed in our previous work~\cite{Li:2022vmb}, the time-reversal symmetry makes the coefficients dissipative, so that T-even non-dissipative coefficients should be zero. However, it should be noted that the above conclusion is stronger than the requirement of T-symmetry. Based on symmetry, there should be a term with a non-dissipative T-even coefficient.
\begin{align}
	\label{eq_p-oddc}
	\sT^{\a\b}_{\rm nd}=c_{\rm nd} u^{\langle\a} \pd^{\beta\rangle}_{\perp}\beta,\; \text{P symmetry makes}\;c_{\rm nd}=0,
\end{align}
while $c_{\rm nd}$ is still zero here. As the answer indicated in Eq.~(\ref{eq_p-oddc}), the coefficient should be ``parity-odd"; otherwise, it is zero. For a system without net axial charge, $c_{\rm nd}$ vanishes. However, for a medium with local axial charge density, as shown in~\cite{Liu:2019krs}, we might see a non-vanishing coefficient (locally).
\subsection{Dissipative contribution}
The dissipative contribution is one of the key discoveries discussed in detail in our previous work~\cite{Li:2022vmb}, where we first established the fluctuation-dissipation relation for spin alignment (noting that spin alignment is defined to be uneven spin fluctuations)
Meanwhile, the phenomenological effects of this contribution are estimated to be significant, potentially generating a $p_T$ and centrality dependence that exhibits similar trends to those observed in experiments.

However, to avoid affecting the publication of our paper~\cite{Li:2022vmb}, we will complete the section by briefly re-deriving the two essential formulas for the dissipative part of the tensor polarization, which can be obtained by replacing $O^a$ with $V^\mu(t, \vx - \vy/2) V^\nu(t, \vx + \vy/2)$ in Eq.~(\ref{eq_od-final}).
Then, we will perform the Wigner transform of $V^\mu V^\nu$ and express $G_R$ in the Fourier space of $\vq$. With $\vq$ taken to zero, the leading gradient contribution is
\begin{align}
	\label{eq_tens_dis_1}
	\sT_{\rm d}^{\a\b}(x,\vp)	=&-\frac{2\ve_{u}}{\beta} \int d^{3}\vx' \pd_\l\beta_\g(t,\vx') \int \frac{d^3\vq}{(2\pi)^3}e^{i\vq\cdot(\vx-\vx')}\no\\
	&\times\lim_{\vq\rightarrow0}^{\o\rightarrow0}\frac{\pd}{\pd\o}\text{Im}G_{R}^{\langle\a\b\rangle\l\g}(\o,\vq,\vp)
\end{align}
Note that higher orders of $\vq$ are neglected, so the only $\vq$ dependence comes from $e^{i\vq \cdot (\vx - \vx')}$. Integrating over $\vq$ and $x'$ will create a $\delta(\vx - \vx')$, which makes $\beta(t, \vx')$ equal to $\beta(t, \vx)$. In the end, the theory becomes local as
\begin{align}
	\label{eq_tens_dis_2}
	\sT_{\rm d}^{\a\b}	(x,\vp)=&-\frac{2\ve_u}{\beta}  \lim^{\o\rightarrow0}_{\vq\rightarrow0}\frac{\pd}{\pd\o}\text{Im}G^{\langle\a\b\rangle\l\g}(\o,\vq,\vp)\pd_\l\beta_\g(t,\vx)
\end{align}
where the correlation function in Eq.~(\ref{eq_def_Gr}) is repeated here for convenience:
\begin{align}
	\label{eq}
	&G^{\mu\nu\l\g}_R(t-t',\vx,\vx',\vy)\equiv\int \frac{d\omega}{2\pi} \frac{d^3\vq}{(2\pi)^3} \frac{d^3\vp}{(2\pi)^3}e^{-i\omega\cdot(t-t')}
	\no\\
	&\hspace{3cm}\times e^{i\vq\cdot(\vx-\vx')}e^{i\vp\cdot\vy}G^{\mu\nu \l\g}_{R}(\omega,\vq,\vp)
	\no\\
	&=(-i)\theta(t-t')\langle [V^\mu(t,\vx_{-}) V^{\nu}(t,\vx_{+}),T^{\l\g}(t',\vx')]\rangle
\end{align}
This is correct to all orders (of interactions). With hadronic basis field theory, we can directly perform non-perturbative calculations of mesons to determine the correlation function. For QCD, some additional effort is required to extract the bound vector meson states from all states in the vector current-current correlations. We might explore this possibility further in the future. In our previous work~\cite{Li:2022vmb}, we explored this dissipative contribution to spin alignment in the simplest scenario, using the leading dressed/skeleton order in the skeleton expansions~\cite{PhysRev.118.1417} (see Sec.~\ref{sec_skeleton}), where interactions between the vector meson and the medium are all encoded in the self-energies and are independent of the details of the interactions. 



\subsection{Skeleton expansion, non-perturbative methods}
\label{sec_skeleton}
In this section, we briefly review the skeleton expansion scheme, which is typically employed when discussing perturbation to all orders. For the skeleton expansion, rather than expanding using bare propagators, it expands the correlation function in terms of full/dressed propagators and full/dressed vertices. The scheme was first proposed by Dyson~\cite{Dyson:1949ha}. Additionally, Luttinger-Ward's work~\cite{PhysRev.118.1417} and Srednicki's book~\cite{Srednicki:2007qs} are useful references for this expansion. It is the underlying expansion scheme for many non-perturbative approaches, such as DSE~\cite{Roberts:2000aa} and FRG~\cite{Fu:2019hdw}.

To make the problem more concrete, we will use the quark-meson model as an example, for which the Lagrangian density is
\begin{align}
	\mathcal{L}=\bar{\psi}(i\slashed{D}-m)\psi-\frac{1}{4}F^{\m\n}F_{\m\n}+\frac{1}{2}m^2 V^\m V_\m
\end{align}
The covariant derivative is defined as $D_{\mu} = \pd_{\mu} - i g_{V} V_{\mu}$. Using a similar procedure as discussed in Sec.~\ref{sec_current}, the Belinfante form of the energy-momentum tensor can be obtained
\begin{align}
	\label{eq_Tmn_QM}
	T^{\mu\nu}=& T^{\mu\nu}_{V}+T^{\mu\nu}_{\p}+T^{\mu\nu}_I\no\\
	=&-\tensor{F}{^{\mu}^{\alpha}}\tensor{F}{^{\nu}_{\alpha}}+m^2V^{\mu}V^{\nu}-\eta^{\mu\nu}(-\frac{1}{4}F^{\alpha\beta}F_{\alpha\beta}\no\\
	&+\frac{1}{2}m^2V^\a V_\a)+\frac{1}{2}\bar{\psi}i\gamma^{(\m} \overset{\leftrightarrow\;\;}{\pd^{\n)}}\psi
	+g_V\bar{\psi}\gamma^{(\m} V^{\n)}\psi\,, 
\end{align}
which is a summation of the free spin-1/2  $T^{\mu\nu}_{\psi}$ in Sec.~\ref{sec_spinhalf} and spin-1 $T^{\mu\nu}_{V}$ in Sec.~\ref{sec_spinoneten}, with ``interaction" terms $T^{\mu\nu}_I = g_V \bar{\psi} \gamma^{(\mu} V^{\nu)} \psi$.

\textit{Spin-1/2 Case--}With this $T^{\mu\nu}$ in Eq.~(\ref{eq_Tmn_QM}), for spin-1/2, the correlation functions defined in Eq.~(\ref{eq_j5-half-gen-1}) will be related to the following correlation functions and their derivatives:
\begin{align}
	\label{eq_Gskeleton_half}
&G(1,1',2,2')=-\<T_\tau[\bar{\psi}(1)\psi(1')\bar{\psi}(2)\psi(2')]\>\no\\
&G^\l(1,1',2,2',3)=-\<T_\t[\bar{\psi}(1)\psi(1')\bar{\psi}(2)\psi(2')V^\l(3)]\>\no\\
&G^{\l\g}(1,1',2,2')=-\<T_\t[\bar{\psi}(1)\psi(1')V^\l(2)V^\g(2')] \>
\end{align}
The $G$, $G^\l$ and $G^{\l\g}$ are related to the $\< \bar{\psi}\psi T_{\psi}^{\a\b}\>$, $\< \bar{\psi}\psi T_I^{\a\b}\>$ and  $\< \bar{\psi}\psi T_{V}^{\a\b}\>$ terms within the full $\< \bar{\psi}\psi T^{\a\b}\>$ respectively.


\begin{figure} [h]
	\centering
	\includegraphics[width=0.99\columnwidth]{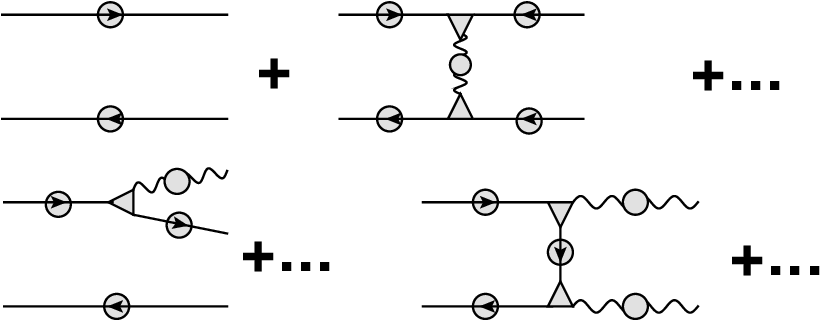}
	\caption{Examples of the diagrams for the correlation functions for spin-1/2 particles. First line is for $G$. The diagrams on the left or right of second line is for $G^\mu$ or $G^{\mu\nu}$.
	}
	\label{fig_half}
\end{figure}


Examples of the diagrams for these correlation functions are shown in Fig.\ref{fig_half}. Following the hierarchy of this diagrammatic expansion, the calculation performed for the vector polarization of spin-1/2 corresponds to the closed loop with the first diagram of Fig.\ref{fig_half}, which is classified as leading order and only related with the ``free" spin-1/2 $T^{\a\b}_{\p}$ through $\< \bar{\psi}\psi T_{\psi}^{\a\b}\>$.
However, in that calculation, we also approximate the ``full" propagator by the ``free" one, which further simplifies the calculation while retaining the main physics.


\textit{Spin-1 Case--}
Similarly, for spin-1 particles, with $T^{\mu\nu}$ discussed in Eq.~(\ref{eq_Tmn_QM}), the correlation function in Eq.~(\ref{eq_def_Gr_E}) will be related to the following correlation functions and their derivatives:
\begin{align}
	\label{eq_Gskeleton_one}
	&G^{\mu\nu\l\g}(1,1',2,2')=-\<T_\t[V^\m(1)V^\n(1')V^{\l}(2)V^{\g}(2')]\>\no\\
	&G^{\mu\nu\l}(1,1',2,2',3)=-\< T_\t[V^\m(1)V^\n(1')\bar{\psi}(2)\psi(2')V^\l(3)]\>\no\\
	&G^{\mu\nu}(1,1',2,2')=-\<T_\t[ V^\mu(1)V^\nu(1') \bar{\psi}(2)\psi(2')]\>
\end{align}
The $G^{\mu\nu\l\g}$, $G^{{\mu\nu\l}}$ and $G^{\mu\nu}$ are related to the $\< V^\m V^{\n} T_V^{\a\b}\>$, $\< V^\m V^{\n} T_I^{\a\b}\>$ and  $\< V^\m V^{\n}  T_{\p}^{\a\b}\>$ terms within the full $\< V^\m V^{\n} T^{\a\b}\>$ respectively

\begin{figure} [h]
	\centering
	\includegraphics[width=0.99\columnwidth]{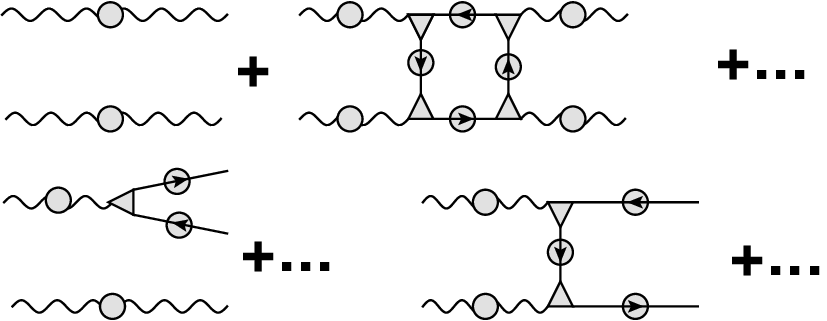}
	\caption{Examples of the diagrams for the correlation functions for spin-1 particles. First line is for $G^{\mu\nu\l\g}$. The diagrams on the left or right of second line is for $G^{{\mu\nu\l}}$ or $G^{\mu\nu}$.
	}
	\label{fig_one}
\end{figure}
\begin{figure} [ht]
	\centering
			\includegraphics[width=0.99\columnwidth]{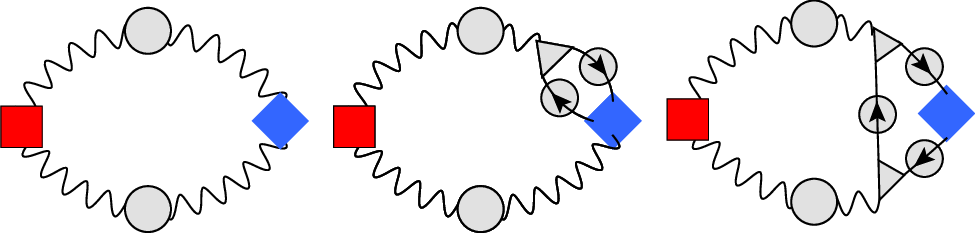}
			\caption{The examples on loop structures for Wigner functions, the red box on the left represent $W^{\m\n}$ and the blue box on the right represent the $\pd_\g u_\l$.  
				}
			\label{fig_close}
\end{figure}

Examples of the diagrams for those correlation functions are shown in Fig.~\ref{fig_one}. If we close the open lines of Fig.~\ref{fig_one} into a loop, which is what is finally evaluated, we obtain the loop structures shown in Fig.~\ref{fig_close}.
Both Fig.~\ref{fig_one} and Fig.~\ref{fig_close} suggest that the leading diagram is the first one, and the leading diagram comes from the term $\< V^\m V^{\n} T_V^{\l\g}\>$, which is the correlation between the Wigner function of the vector boson and the ``free" vector $T_V^{\l\g}$. However, this does not mean they are free from interactions, since interactions between the vector meson and the medium can be encoded in the bubbles/self-energies on the propagator. The interactions encoded in the self-energies contribute to the  spin-alignment discussed in our previous work~\cite{Li:2022vmb}.
As inputs to the formalism are the full propagator/spectral function, with external non-perturbative information on it, we can study some part of non-perturbative effects on spin alignment.

\textit{High Orders, Non-Perturbative Methods--}
The concrete calculations presented so far in this work represent the leading non-zero effects. However, within the expansion scheme discussed in this section, many improvements can be made in the future.

For example, even if we still keep the discussion within the leading skeleton order, we can improve the self-energies through microscopic calculations or through external phenomenological inputs. To go beyond, we can perform two or more loop calculations to include vertex corrections and other more sophisticated considerations~\cite{PhysRev.118.1417,Baym:1961zz}. On the other hand, we have many non-perturbative theoretical tools for the evaluation of the correlation functions, especially in Euclidean time. For example, the vertex structure and the self-energies can be evaluated using the DSE~\cite{Roberts:2000aa} or FRG~\cite{Fu:2019hdw} methods mentioned previously. Meanwhile, four-point functions can be resolved by the Bethe-Salpeter equation or its reduced 3D $T$-matrix equations~\cite{Liu:2016ysz,Liu:2017qah}. We might even employ lattice QFT to directly calculate these correlation functions from first principles, as discussed in Sec.~\ref{sec_conti}.

\section{Summary and perspective}
\label{sec_sum}
In this work, we have discussed the theoretical foundation and essential techniques of our Zubarev response approach for calculating the vector and tensor polarizations of particles in a locally equilibrated hydrodynamic medium. In particular, we have systematically explained how to incorporate familiar correlation functions and diagrammatic techniques within the Zubarev response approach.

For the concrete application of the Zubarev response approach, we list several outcomes and relevant subtleties that warrant some attention:
\begin{adjustwidth}{0.5cm}{}
	(1) We have reproduced the recently discovered local vector polarization formula for spin-1/2 particles~\cite{Liu:2021uhn} using the Zubarev response approach, including the shear-induced polarization contributions.
	
	\noindent(2) The local vector polarization for spin-1 particles has been derived, and it follows the same form as the spin-1/2 case, with an additional factor of 4/3.
	
	\noindent(3) We explicitly prove that non-dissipative part of the first-order gradient contribution to spin alignment vanishes.
	
	\noindent(4) We present an alternative derivation for Zubarev response approach. 
	
	\noindent(5) We discuss the subtleties and relations between slow and fast modes, the slow-fast limits.
	
	\noindent(6) We illustrate the subtleties and relations between different definitions of density matrices and how to make the final operator average covariant for various definitions.
	
	\noindent(7) We explain the skeleton expansion implicitly used in previous calculations and show how to include higher-order terms, and how to relate to non-perturbative methods, including lattice QFT.
\end{adjustwidth}

In the future, a meaningful extension of this work is to go beyond the linear order of gradients to include non-linear response terms, which have been explored in other methods~\cite{Sheng:2024pbw,Zhang:2024mhs,Yang:2024fkn}. Also, as a diagrammatic approach, we can explore higher-order (interaction) effects using typical perturbation theory. More ambitiously, since polarization quantities can have non-perturbative definitions, we can calculate various polarizations using non-perturbative methods such as DSE, FRG, and lattice QFT, which may ultimately lead to a much deeper understanding of the polarization phenomena in heavy-ion collisions and help address the ongoing puzzles in the field.

\acknowledgments  
We thank Yi Yin, Feng Li for valuable discussions.
This research is supported by NSFC No. 12205090.\\





\bibliography{ref}

\end{document}